\documentclass[aps,prd,superscriptaddress,
groupaddress,twocolumn,nofootinbib]{revtex4-1}

\usepackage{amsmath}
\usepackage{amssymb}
\usepackage{graphicx}
\usepackage{mathrsfs}
\usepackage{hyperref}
\usepackage{braket}

\def\eref#1{(\ref{eq:#1})}
\def\roundbr#1{\left(#1\right)}
\def\squarebr#1{\left[#1\right]}
\def\curlybr#1{\left\{#1\right\}}
\def\angbr#1{\left<#1\right>}
\def\roundprod#1#2{\roundbr{#1\left|#2}\right.}
\def\angprod#1#2{\angbr{#1\left|#2}\right.}
\def\del#1#2{\frac{\partial #1}{\partial #2}} 
\def\D{\mathrm{d}}
\def\e{\mathrm{e}}
\def\I{\mathrm{i}}
\def\Ob{\mathcal{O}}
\def\n{\mathbf{n}}
\def\q{\mathbf{q}}
\def\Ncal{\mathcal{N}}
\def\x{\mathbf{x}}
\def\K{\mathbf{k}}
\def\He{\hat{H}_{\text{eff}}}

\usepackage[normalem]{ulem}

\usepackage{xcolor}

\begin{document}

\title{Unitarity of quantum-gravitational corrections to primordial fluctuations in the Born-Oppenheimer approach}
 
\author{Leonardo Chataignier}
\email{lcmr@thp.uni-koeln.de}
\affiliation{Institut f\"ur Theoretische Physik, Universit\"{a}t zu
   K\"{o}ln, Z\"{u}lpicher Stra\ss e 77, 50937 K\"{o}ln, Germany} 
   
\author{Manuel Kr\"{a}mer}
\affiliation{Institute for Theoretical Physics, KU Leuven,
 Celestijnenlaan 200D, 3001 Leuven, Belgium} 
 
\date{\today}

\begin{abstract}
We revisit the calculation of quantum-gravitational corrections to the power spectra of scalar and tensor perturbations in the Born-Oppenheimer approach to quantum gravity. We focus on the issue of the definition of the inner product of the theory and the unitarity of the corrections to the dynamics of the cosmological perturbations. We argue that the correction terms are unitary, provided the inner product is defined in a suitable way, which can be related to a notion of gauge fixing the time variable and the use of conditional probabilities in quantum cosmology. We compare the corrections obtained within this framework to earlier results in the literature and we conclude with some remarks on the physical interpretation of the correction terms.
\end{abstract}

\maketitle

\section{Introduction}
Any candidate theory of quantum gravity must address the issue of producing testable predictions. It is reasonable to expect that the early Universe may be an adequate testing ground for quantum gravity and, indeed, a lot of effort~\cite{BKK-0-0,BKK-0-1,BKK-1,BKK-2,Mariam,Brizuela:moment,BKK-3,Stein1,Stein2,Bologna:2013,Bologna:2014,Bologna:2016,Bologna:2018,Bologna:2020,Ashtekar:2006,Barrau:2018,Martineau:2018,MalMiro:2020} has been devoted to calculating primordial quantum-gravitational effects. In particular, many lines of inquiry look for quantum-gravitational corrections to the Cosmic Microwave Background (CMB) anisotropy spectrum. Within the inflationary paradigm, this is justified by considering that quantum fluctuations of the metric and the inflaton field give rise to the CMB anisotropies and the conditions for structure formation. Although effects of quantum gravity are expected to become sizable at energies close to the Planck scale, it is possible that some corrections are already relevant at the high energies present during the inflationary phase.

In the present article, we revisit the calculation of quantum-gravitational corrections to the dynamics in the early Universe using the conservative approach based on the canonical quantization of general relativity in metric variables, which leads to the so-called Wheeler-DeWitt (WDW) constraint~\cite{DeWitt:1967,Kuchar:1991,Isham:1992}. This might not be the most fundamental approach, but it has a straightforward classical limit and it is expected to be valid at least within the energy scales that we consider~\cite{Kiefer:book}. Moreover, quantum field theory (QFT) in curved spacetimes can be derived from the WDW constraint if one uses a weak-coupling expansion~\cite{LapRuba:1979,Banks:1984,Halliwell:1984,Brout:1987-4,PadSingh:1990-1,PadSingh:1990-2,Kiefer:1993-2,Parentani:1997-2} that resembles the Born-Oppenheimer (BO) approximation frequently used in nuclear and molecular physics~\cite{BO:1927,Cederbaum:2008,Abedi:2010,Arce:2012}. Previous research has shown that corrections to the dynamics of quantum fields on a given background can be derived using this BO approach~\cite{Singh:1990,Kiefer:1991}. Nevertheless, some of these corrections were thought to violate unitarity and were thus neglected~\cite{Kiefer:1991,BKK-0-0,BKK-0-1,BKK-1,BKK-2,BKK-3,Stein2}. In~\cite{Bertoni:1996,Bologna:2017,Kiefer:2018}, it was then argued that unitarity is preserved if one includes the effect of ``backreaction'' and ``non-adiabatic'' contributions in the BO weak-coupling expansion. The issue was revisited in~\cite{Chataig:2019-1}, where it was emphasized that the backreaction terms are a priori ambiguous and that the unitarity of the theory rests on a choice of inner product related to the choice of time variable in quantum gravity. The work of~\cite{Chataig:2019-1} was, however, limited by the use of the Klein-Gordon inner product, which is not positive-definite.

Our goal in this article is to assess whether the quantum-gravitational corrections to the dynamics in the BO approach are, in fact, unitary and how the definition of a positive-definite inner product is related to a notion of gauge fixing the time variable in quantum cosmology. Our work is a continuation of~\cite{Chataig:2019-1} and is inspired by recent developments in relational approaches to quantum dynamics~\cite{Chataig:2019-2,Hoehn:2018-1,Hoehn:2018-2,Hoehn:2019,Hoehn:Trinity,Chataig:2020,Hoehn:2020}. For simplicity, we will examine the case of cosmological perturbations on a de Sitter background. The application of the BO approach to general slow-roll models was considered in~\cite{BKK-2}, and we will conclude with some comments on the implications of our results on unitarity to these more general cases. 

The article is organized as follows. In Sec.~\ref{sec:DeSitter}, we set up the classical and quantum theories of cosmological perturbations on a (quantum) Friedmann-Lama\^{\i}tre-Robertson-Walker (FLRW) background, specializing to the de Sitter case. Based on a notion of gauge fixing the time variable, we discuss the definition of a positive-definite inner product for both the background and perturbation variables, to which we refer as the gauge-fixed inner product. We also argue that the physical predictions associated with this inner product should be interpreted in terms of conditional probabilities. In Sec.~\ref{sec:expansion}, we define the BO weak-coupling expansion and we argue that the two versions of the BO approach used in~\cite{Singh:1990,Kiefer:1991,BKK-0-0,BKK-0-1,BKK-1,BKK-2,BKK-3,Stein1,Stein2} and~\cite{Bologna:2013,Bologna:2014,Bologna:2016,Bologna:2018,Bertoni:1996,Bologna:2017} are equivalent. Furthermore, we show that the dynamics is unitary with respect to the gauge-fixed inner product. In Sec.~\ref{sec:corrections}, we use the formalism here developed to revisit the calculation of the corrections to primordial power spectra in the BO approach. We include certain terms which were previously neglected, as they were considered to violate unitarity in~\cite{BKK-0-0,BKK-0-1,BKK-1,BKK-2,BKK-3,Stein2}. Consequently, our results differ to a certain extent from those reported in those references. In particular, we comment on the appearance of a late-time logarithmic term which may jeopardize the validity of perturbation theory. In Sec.~\ref{sec:conclusions}, we summarize our results and present our conclusions. Summation over repeated indices is implied and we work in units in which $c = \hbar = 1$. Spacetime is assumed to be four-dimensional and the metric signature is mostly pluses.

\section{\label{sec:DeSitter}Cosmological perturbations on a FLRW background}
We consider a flat FLRW universe with compact spatial topology as the background on which cosmological perturbations are defined. This is justified due to the fact that the Universe is to a good approximation homogeneous and isotropic at large scales and that, within the inflationary paradigm, spatial curvature contributions are flattened by inflation. Here, we will review only the basic aspects of the classical theory needed for our analysis of the quantum theory and, in particular, the BO approach. The reader is referred to~\cite{Mukhanov} for further details concerning the theory of cosmological perturbations and to~\cite{Langlois:1994,Malk} for an account of the Hamiltonian formalism for perturbations in general relativity.

\subsection{The classical FLRW background}
We assume the background line element to be
\begin{equation}
\D s^2 = -N^2(t)\D t^2 +a^2(t)\D{\bf x}^2 \,,
\end{equation}
where $N(t)$ is the lapse function associated with the choice of time coordinate $t$ and $a(t)$ is the scale factor of the universe. Under a time reparametrization, the lapse transforms as a density,
\begin{equation}\label{eq:repar-density}
t\mapsto t' \,, \quad N\mapsto N'\frac{\D t'}{\D t} \,.
\end{equation}
In this way, the action for a flat FLRW universe with vanishing cosmological constant and a minimally coupled inflaton field $\phi$ reads~\cite{Kiefer:book}
\begin{equation}
    S = \int_{t_0}^{t_1}\mathrm{d}t\ \mathfrak{L}^3 N\left(-\frac{1}{2\kappa}\frac{a\dot{a}^2}{N^2}+\frac{a^3\dot{\phi}^2}{2N^2}-a^3\mathcal{V}(\phi)\right)
\end{equation}
where $\cdot\equiv\D/ \D t$, $\kappa = 4\pi G/3$ and $\mathfrak{L}^3$ is the volume associated with some arbitrary length scale $\mathfrak{L}$. For convenience, we make the following redefinitions~\cite{BKK-1,BKK-2,Bologna:2014}
\begin{equation}\label{eq:Lfrak-redef}
\begin{aligned}
    a &\mapsto \frac{a}{\mathfrak{L}} \,, \quad N \mapsto \frac{N}{\mathfrak{L}} \,,\\
    t &\mapsto \mathfrak{L}t \,, \quad \x\mapsto\mathfrak{L}\x \,, 
\end{aligned}
\end{equation}
such that the scale factor and the lapse function acquire dimensions of length, while the spacetime coordinates are now dimensionless. The action can then be written as
\begin{equation}
    S = \int_{t_0}^{t_1}\mathrm{d}t\ \left(-\frac{1}{2\kappa}\frac{a\dot{a}^2}{N}+\frac{a^3\dot{\phi}^2}{2N}-Na^3\mathcal{V}(\phi)\right) ,
\end{equation}
or in Hamiltonian form
\begin{equation}\label{eq:flrw-action}
    S = \int_{t_0}^{t_1}\mathrm{d}t\ \left(p_a\dot{a}+p_{\phi}\dot{\phi}-NC\right) ,
\end{equation}
where 
\begin{equation}\label{eq:classical-flrw-C}
    C = -\frac{\kappa}{2a}p_a^2+\frac{1}{2a^3}p_{\phi}^2+a^3\mathcal{V}(\phi) \,.
\end{equation}
From~\eref{classical-flrw-C}, we see that the configuration space is endowed with the metric
\begin{equation}\label{eq:minimetric-a-phi}
\mathcal{G} := \mathrm{diag}\roundbr{-\frac{a}{\kappa},a^3} ,
\end{equation}
such that the kinetic term may be written as $\frac12\mathcal{G}^{ij}p_ip_j$ ($i,j = a,\phi$), where $\mathcal{G}^{ij}$ are the coefficients of the inverse of~\eref{minimetric-a-phi}.

Due to the assumption of homogeneity and isotropy, the action has been reduced to that of a mechanical theory.\footnote{This is due to the fact that this model obeys the ``symmetric criticality principle''; i.e., the critical points of the action~\eref{flrw-action} correspond to critical points of the full Einstein-Hilbert action. This is an instance of the symmetry reduction procedure (for the case of homogeneity and isotropy). The ``reduction'' consists in the construction of invariant fields and their corresponding field equations for a given symmetry group. In general, however, symmetric criticality is not satisfied; i.e., the reduction of the action does not yield the same results as the reduction of the field equations. See~\cite{Kiefer:book,Fels:2002} for further details.} Such mechanical (toy) theories of cosmology are often called minisuperspace models and they can be seen as generalizations of the theory of a relativistic particle. The $t$-manifold corresponds to the the ``worldline'', whereas the lapse corresponds to the ``einbein''. A ``change of frame'' for the einbein is described by the redefinitions
\begin{equation}\label{eq:einbein-frame}
N(t) = \tilde{N}(t)\Omega(t) \,, \quad C = \frac{1}{\Omega(t)}\tilde{C} \,,
\end{equation}
which leave the Hamiltonian $NC$ invariant. The function $\Omega(t)$ is arbitrary but non-vanishing. It may depend on the configuration space variables. If $\Omega(t)>0$, the transformation~\eref{einbein-frame} induces a conformal transformation in minisuperspace
\begin{equation}\label{eq:mini-conformal}
\mathcal{G}_{ij} = \Omega(t)\tilde{\mathcal{G}}_{ij} \,, \quad a^3\mathcal{V}(\phi) = \frac{a^3\tilde{\mathcal{V}}(a,\phi)}{\Omega(t)} \,,
\end{equation}
which will be of relevance later.

As the lapse function $N(t)$ appears in~\eref{flrw-action} as a Lagrange multiplier, we obtain the Hamiltonian initial-value constraint $C = 0$, which is a consequence of the time-reparametrization invariance of the theory. Indeed, the time evolution of a phase-space function $f(a,\phi,p_a,p_{\phi})$ is given by
\begin{equation}
    \dot{f} = \{f,NC\} \approx N\{f,C\} \,,
\end{equation}
where ``$\{\cdot,\cdot\}$'' is the Poisson bracket and ``$\approx$'' is Dirac's weak equality sign~\cite{Dirac:1958-1}, which denotes identities that hold on the constraint surface defined by $C = 0$ (i.e., on-shell identities). Thus, due to~\eref{repar-density} and~\eref{einbein-frame}, the equations of motion have the same form for any choice of time parameter $t$.

We refer to a choice of $t$ [or a choice of $N(t)$] as a gauge fixing of the time variable. Note that one can fix $N(t)$ regardless of the choice of einbein frame. In general, if we use the level sets of a phase-space function $\chi(a,\phi,p_a,p_{\phi})$ to define the time variable, the lapse is determined via the formula
\begin{equation}\label{eq:FP-det}
\frac{1}{N} \approx \Delta_{\chi} := \{\chi,C\} \,,
\end{equation}
such that $\dot{\chi}\approx 1$.\footnote{We can also write the gauge fixing condition as an extra constraint with an explicit time dependence, $\tilde{\chi}:=\chi-t$. Then, the lapse is determined by requiring that this constraint is preserved by evolution, $\dot{\tilde{\chi}}\approx0$~\cite{Barvinsky}.} The quantity $\Delta_{\chi}$ is the Faddeev-Popov ``determinant''\footnote{More generally, one can work with a reparametrization-invariant extension of~\eref{FP-det}, which can be defined by the formula $|\Delta_{\chi}|^{-1} = \int\D t\ \delta(\chi-t)g$, where $g$ is a function that restricts the integration to a region where the gauge condition is admissible~\cite{Chataig:2019-2,Chataig:2020,HT:book}. Such an integral formula is frequently used in the Faddeev-Popov gauge-fixing procedure for path integrals. However, we shall not use it here because~\eref{FP-det} is sufficient for our purposes.} associated with the canonical ``gauge condition'' $\chi$~\cite{HT:book}. The gauge fixing is admissible in regions of phase space where $\Delta_{\chi}\neq0$. If the einbein frame is changed according to~\eref{einbein-frame} for a fixed gauge condition $\chi$, the function $\Delta_{\chi}$ transforms accordingly to
\begin{equation}\label{eq:FP-einbein-change}
\tilde{\Delta}_{\chi} := \{\chi,\tilde{C}\} \approx \Omega(t)\Delta_{\chi} \,.
\end{equation}
Certain gauge fixings of the time variable can be expressed as functions in configuration space, provided one makes use of a solution of the Hamilton-Jacobi equation, such as the on-shell action. The Hamilton-Jacobi equation that corresponds to~\eref{classical-flrw-C} is
\begin{equation}\label{eq:HJ-flrw}
-\frac{\kappa}{2a}\roundbr{\del{S}{a}}^2+\frac{1}{2a^3}\roundbr{\del{S}{\phi}}^2+a^3\mathcal{V}(\phi) = 0 \,.
\end{equation}
By setting
\begin{align*}
p_a = \del{S}{a} \,, \quad p_{\phi} = \del{S}{\phi} \,,
\end{align*}
we can express the on-shell time evolution of a dynamical quantity $\chi$ as
\begin{equation}\label{eq:config-time-function-flrw-0}
\frac{1}{N}\dot{\chi} = -\frac{\kappa}{a}\del{S}{a}\del{\chi}{a}+\frac{1}{a^3}\del{S}{\phi}\del{\chi}{\phi} \,.
\end{equation}
The quantity $\chi(a,\phi)$ is a configuration-space time function if $\dot{\chi} = 1$ for a given choice of $N$. Thus, the configuration-space time function must be a solution of [cf.~\eref{config-time-function-flrw-0}] (see also~\cite{Kuchar:1991,Isham:1992})
\begin{equation}\label{eq:config-time-function-flrw}
-\frac{\kappa}{a}\del{S}{a}\del{\chi}{a}+\frac{1}{a^3}\del{S}{\phi}\del{\chi}{\phi} = \frac{1}{N} \,.
\end{equation}
For simplicity, let us now consider the de Sitter (``no-roll'') limit of the theory. In this limit, the scalar field is constant. When one inserts $\phi =$ const. into the equations of motion for the inflaton,
\begin{equation}
    \dot{\phi} \approx \frac{N}{a^3}p_{\phi} \,, \quad \dot{p}_{\phi} \approx -Na^3\frac{\partial\mathcal{V}}{\partial\phi} \,, \label{eq:eom-inflation-flrw}
\end{equation}
one obtains the conditions
\begin{equation}\label{eq:no-roll-flrw}
p_{\phi} = \del{\mathcal{V}}{\phi} = 0 \,. 
\end{equation}
In this way, the inflaton potential is independent of $\phi$ and we define it to be~\cite{BKK-1}
\begin{equation}\label{eq:DeSitterCase-V}
    \mathcal{V}(\phi) = \frac{H_0^2}{2\kappa} \,,
\end{equation}
where $H_0$ is the Hubble parameter for a de Sitter universe. The constraint~\eref{classical-flrw-C} can then be solved for $p_a$ to yield
\begin{equation}\label{eq:pa-de-sitter}
p_a = -\frac{\sigma H_0}{\kappa} a^2 \,, \quad (\sigma=\pm1) \,.
\end{equation}
The choice of $\sigma = 1$ corresponds to an expanding universe, whereas $\sigma = -1$ corresponds to a contracting one. This discrete multiplicity is analogous to the positive- and negative- frequency sectors of a relativistic particle~\cite{Chataig:2019-2}. Using $\dot{\phi} = 0$ and~\eref{pa-de-sitter} in~\eref{flrw-action}, we find the on-shell action
\begin{equation}\label{eq:deSitter-on-shell-action}
S_{C = 0} = \int_{a_0}^{a_1}p_a\D a = -\frac{\sigma H_0}{3\kappa} (a_1^3-a_0^3) \,,
\end{equation}
which is a solution to the Hamilton-Jacobi equation~\eref{HJ-flrw}. According to the discussion preceding~\eref{config-time-function-flrw}, a general choice of time coordinate $\eta$ can be expressed in terms of the configuration space variables (in this case, the scale factor) through the following equation:
\begin{equation}\label{eq:time-function-de-sitter}
    -\frac{\kappa}{a_1}\frac{\partial S_{C=0}}{\partial a_1}\frac{\partial\eta}{\partial a_1} = \frac{1}{N} \,.
\end{equation}
We will be interested in conformal time, for which $N = a_1$. This yields
\begin{equation}\label{eq:conformal-time}
    \eta(a) = -\frac{\sigma}{H_0 a}
\end{equation}
as a solution to~\eref{time-function-de-sitter} (we dropped the subscript from $a_1$). For consistency, let us verify that, instead of deriving~\eref{conformal-time} from the choice $N = a$, one can follow the opposite path and calculate the on-shell lapse function from the choice of time~\eref{conformal-time}. The Faddeev-Popov determinant reads [cf.~\eref{FP-det}]
\begin{equation}\label{eq:FP-det-eta}
\Delta_{\eta} = \{\eta(a),C\} = -\frac{\sigma\kappa}{H_0 a^3}p_a \,.
\end{equation}
Using~\eref{pa-de-sitter}, which holds when $C = 0$, we can rewrite~\eref{FP-det-eta} as
\begin{equation}\label{eq:FP-det-eta-on-shell}
\Delta_{\eta} \approx \frac{1}{a} = \frac{1}{N} \,.
\end{equation}
Hence, $N = a$, as it should be. Finally, for later reference, we change the einbein frame such that $N = \tilde{N}a$, and subsequently we use~\eref{FP-einbein-change} and~\eref{conformal-time} to rewrite~\eref{FP-det-eta} in terms of conformal time,
\begin{equation}\label{eq:FP-det-eta-eta}
\tilde{\Delta}_{\eta} = -\kappa H_0^2\eta^4p_{\eta} \,,
\end{equation}
where $p_{\eta}$ is the canonical momentum conjugate to $\eta$. From~\eref{FP-det-eta-on-shell}, we obtain $\tilde{\Delta}_{\eta} \approx 1$, as it should be.

\subsection{\label{sec:quantum-flrw}The quantum FLRW background}
In the quantum theory, one must account for the constraint~\eref{classical-flrw-C} in some way. In most approaches to canonical quantum gravity and, in particular, to quantum cosmological toy models, the quantum analogue of~\eref{classical-flrw-C} is the quantum constraint equation~\cite{DeWitt:1967}
\begin{equation}\label{eq:background-WDW-0}
\hat{C}\roundbr{a,\phi,-\I\del{}{a},-\I\del{}{\phi}}\Psi(a,\phi) = 0 \,,
\end{equation}
which is also known as the WDW equation~\cite{Kiefer:book}. This is also the approach we adopt here.\footnote{See, however, the approach of St\"{u}ckelberg and its later incarnations~\cite{Stueckelberg:1942, Carlini-Greensite,Gryb}, in which the quantum constraint is \emph{not} imposed and one allows quantum states that are ``off shell'' according to the terminology we adopt. The main reasoning behind this approach is that the classical value of the constraint should be identified with a constant of motion, the value of which is determined by the initial conditions. This was first proposed for the free relativistic particle~\cite{Stueckelberg:1942}, where the value of the particle's mass was taken to depend on its initial position and momentum. In this way, instead of imposing the WDW equation in the quantum theory, one works with the usual Schr\"{o}dinger equation, and a specific value for the constraint can be recovered in correlation functions computed from certain peaked states (e.g., $\braket{\hat{C}} = 0$).} More precisely, we consider the equation $\Omega_0(\alpha,\phi)\hat{C}\Psi(\alpha,\phi) = 0$, where $\Omega_0$ is a positive configuration-space function. We take a choice of $\Omega_0$ to be the quantum analogue of a choice of einbein frame [cf.~\eref{einbein-frame}]. We also choose the Laplace-Beltrami factor ordering for the kinetic term in~\eref{classical-flrw-C}, 
\begin{equation}\label{eq:WDW-flrw-0}
\Omega_0\squarebr{\frac{\kappa}{2a^2}\del{}{a}\roundbr{a\del{}{a}}-\frac{1}{2a^3}\del{^2}{\phi^2}+a^3\mathcal{V}(\phi)}\Psi(a,\phi) = 0 \,,
\end{equation}
such that the quantum constraint is invariant under coordinate transformations in configuration space. Moreover, it is straightforward to verify that~\eref{WDW-flrw-0} has the same form if one performs the transformation $\Omega_0 = \tilde{\Omega}_0\Omega$ together with~\eref{mini-conformal}.\footnote{The fact that the constraint~\eref{background-WDW-0} is invariant under such conformal transformations is a consequence of the fact that the background configuration space is two-dimensional. For higher-dimensional cases, one can adopt a more general conformal factor ordering~\cite{Misner:1972,Halliwell:conformal}, in which the WDW equation~\eref{background-WDW-0} is conformally covariant.} It is also convenient to adopt the variable~\cite{BKK-1}
\begin{equation}
\alpha = \log\roundbr{\frac{a}{a_0}} ,
\end{equation}
where $a_0$ is some reference scale factor, because $\alpha$ takes values over the real line, whereas $a >0$. In this way, the background quantum constraint reads
\begin{equation}\label{eq:WDW-flrw}
\Omega_0\frac{\e^{-3\alpha}}{a_0^3}\squarebr{\frac{\kappa}{2}\del{^2}{\alpha^2}-\frac{1}{2}\del{^2}{\phi^2}+a_0^6\e^{6\alpha}\mathcal{V}(\phi)}\Psi(\alpha,\phi) = 0 \,.
\end{equation}

Although we are primarily interested in the quantum theory of perturbations [cf.~Secs.~\ref{sec:pert} and~\ref{sec:master-WDW}], it is worthwhile to comment on the interpretation of the quantum theory of the solutions of~\eref{WDW-flrw}, i.e., the minisuperspace quantum mechanics. It is far from clear whether the traditional Hilbert space structure should be constructed for the on-shell states $\Psi(\alpha,\phi)$. As is well-known, there are two main reasons for this: (1) Whereas in traditional quantum mechanics an external time parameter is available (that roughly corresponds to ``laboratory time''), such a parameter is not present in~\eref{WDW-flrw}, which is a stationary Schr\"{o}dinger equation. The challenge of understanding the quantum dynamics solely from such stationary states is the so-called problem of time~\cite{Kuchar:1991,Isham:1992}. There are various approaches and tentative solutions to this problem~\cite{Anderson:book}. One reasonable point of view is that the absence of an external time is a consequence of time-reparametrization invariance and signals that the evolution should be understood in relational terms (e.g., based on ``intrinsic times''). (2) The lack of an external (or preferred) time parameter in~\eref{WDW-flrw} also calls into question the precise definition of the inner product and the probabilistic interpretation. More precisely, with respect to which time variables (if any) is the (emergence of the) Born rule valid?\footnote{These difficulties may in principle be circumvented in the approach of St\"{u}ckelberg~\cite{Stueckelberg:1942} because there the constraint is not imposed and, therefore, one deals with a time-dependent Schr\"{o}dinger equation instead of a time-independent one [cf.~\eref{WDW-flrw}].}

Despite these conceptual problems, it is possible to define tentative choices of the inner product and Hilbert space for the on-shell states $\Psi(\alpha,\phi)$. Whether any of these choices is realized in nature is, of course, an open problem. In what follows, we are going to briefly examine two possible definitions of the physical inner product.

We first note that the quantum constraint is symmetric with respect to the auxiliary inner product
\begin{equation}\label{eq:auxiliary-IP-flrw}
\angprod{\Psi_{(1)}}{\Psi_{(2)}}:=\int_{\mathbb{R}^2}\!\!\D\alpha\D\phi\ \mu_{\text{kin}}\Psi^{*}_{(1)}(\alpha,\phi)\Psi_{(2)}(\alpha,\phi) \,,
\end{equation}
where the auxiliary (kinematical) measure is
\begin{equation}\label{eq:kin-mu-0}
\mu_{\text{kin}} := \frac{a_0^3\e^{3\alpha}}{\Omega_0(\alpha,\phi)} \equiv \frac{\sqrt{\kappa}}{\Omega_0(\alpha,\phi)}\left|\del{a}{\alpha}\right|\sqrt{|\det\mathcal{G}|} \ ,
\end{equation}
and $\det\mathcal{G}$ is the determinant of~\eref{minimetric-a-phi}. Using~\eref{mini-conformal} for the choice of einbein frame $\Omega = \Omega_0/\sqrt{\kappa}$, we find
\begin{equation}\label{eq:kin-mu}
\mu_{\text{kin}} = \left|\del{a}{\alpha}\right|\sqrt{|\det\tilde{\mathcal{G}}|} \,,
\end{equation}
which may be interpreted as the square root of the determinant of the (conformally transformed) minisuperspace metric with respect to the $(\alpha,\phi)$ coordinates.

The auxiliary inner product~\eref{auxiliary-IP-flrw} is often inadequate because the states $\Psi^{(1,2)}$, which are solutions of~\eref{WDW-flrw}, are stationary states. This implies that the na\"{i}ve expectation values of Heisenberg-picture operators are time-independent, which is another aspect of the problem of time. Moreover, the inner product~\eref{auxiliary-IP-flrw} diverges if zero is in the continuous part of the spectrum of the constraint $\Omega_0\hat{C}$.

One well-known and reasonable choice that regularizes the inner product is the Rieffel induced inner product~\cite{Rieffel:1974,HT-SUSY:1982,Marolf:1995-4}, which can be defined as follows. We assume that the constraint is self-adjoint (or that it is possible to choose a self-adjoint extension) with respect to the auxiliary product~\eref{auxiliary-IP-flrw}, such that we may find a complete orthonormal system of eigenstates $\Psi_{E,\n}(\alpha,\phi)$, which are solutions to the eigenvalue equation
\begin{equation}\label{eq:WDW-flrw-general-eigen}
\Omega_0\frac{\e^{-3\alpha}}{a_0^3}\squarebr{\frac{\kappa}{2}\del{^2}{\alpha^2}-\frac{1}{2}\del{^2}{\phi^2}+a_0^6\e^{6\alpha}\mathcal{V}(\phi)}\Psi_{E,\n} = E\Psi_{E,\n} \,.
\end{equation}
The label $\n$ represents possible degeneracies. Orthonormality is defined with respect to the auxiliary inner product~\eref{auxiliary-IP-flrw},
\begin{equation}\label{eq:auxiliary-IP-flrw-1}
\angprod{\Psi_{E',\n'}}{\Psi_{E,\n}} = \delta(E',E)\delta(\n',\n) \,.
\end{equation}
The symbol $\delta(\cdot,\cdot)$ stands for a Kronecker or Dirac delta, depending on whether the labels are discrete or continuous. We now notice that we are only interested in on-shell states, for which $E = 0$. We can then define from~\eref{auxiliary-IP-flrw-1} the induced inner product for these states as~\cite{HT-SUSY:1982,Chataig:2019-2,Chataig:2020}
\begin{equation}\label{eq:induced-IP-flrw}
\roundprod{\Psi_{E = 0,\n'}}{\Psi_{E = 0,\n}} := \delta(\n',\n)\,,
\end{equation}
which, in contrast to~\eref{auxiliary-IP-flrw-1}, is well-defined even if $\delta(E' = 0, E = 0)$ diverges. The physical (on-shell) Hilbert space is defined as the vector space of superpositions of $\Psi_{E = 0,\n}(\alpha,\phi)$ that are square-integrable with respect to the induced inner product~\eref{induced-IP-flrw}.

The inner product~\eref{induced-IP-flrw} is positive-definite and manifestly independent of any choice of gauge fixing of the time variable. Although this is a desirable property, the connection of~\eref{induced-IP-flrw} with a notion of quantum dynamics is not immediately clear. Indeed, there has been a lot of effort in the literature to describe the quantum dynamics in the physical Hilbert space equipped with~\eref{induced-IP-flrw}. Most approaches focus on the definition of ``relational Dirac observables'', which can be defined as operators that act on the physical Hilbert space (``on-shell operators'') and that represent the values of dynamical quantities with respect to a particular gauge fixing of the time variable.\footnote{More precisely, let us refer to dynamical quantities written with respect to a particular gauge fixing of the time variable simply as ``gauge-fixed'' quantities.  Then, classical relational observables are gauge-invariant extensions of gauge-fixed quantities, i.e., they remain invariant under a change of time coordinate, but encode the dynamics with respect to a particular gauge choice. The precise definition and interpretation of the quantum counterparts of relational observables is a topic of active research~\cite{Chataig:2019-2,Hoehn:2018-1,Hoehn:2018-2,Hoehn:2019,Hoehn:Trinity,Chataig:2020,Hoehn:2020}.} These operators can then be seen as on-shell descriptions of the dynamics in a particular ``time reference frame'' or as (on-shell versions of) a ``gauge-fixed Heisenberg picture''~\cite{Chataig:2019-2,Hoehn:2018-1,Hoehn:2018-2,Hoehn:2019,Hoehn:Trinity,Chataig:2020,Hoehn:2020}.

The definition of quantum relational observables and consequently of the dynamics in the physical Hilbert space equipped with~\eref{induced-IP-flrw} is, however, often complicated. We thus wish to consider another definition of the inner product that is more convenient for our present purposes. In analogy to more familiar gauge theories (understood as Hamiltonian systems with first-class constraints~\cite{HT:book}), we seek a definition of transition amplitudes that takes into account a particular choice of gauge (i.e., a time variable in the present context). Usually, one performs the Faddeev-Popov procedure to regularize the path-integral expression for transition amplitudes or partition functions of gauge theories~\cite{FP-1,FP-2}. This involves the insertion of a delta function of the gauge condition together with the Faddeev-Popov determinant in the path integral. Motivated by this well-known procedure, we define the inner product
\begin{equation}\label{eq:FP-IP-flrw}
\begin{aligned}
&\roundprod{\Psi_{(1)}}{\Psi_{(2)}}\\
&:= \sum_{\sigma}\int_{\mathbb{R}^2}\D\alpha\D\phi\ \left(\hat{\mu}^{\frac12}\Psi_{(1)}^{\sigma}\right)^{*}|J|\delta(\chi(\alpha,\phi)-t)\hat{\mu}^{\frac12}\Psi_{(2)}^{\sigma} \,,
\end{aligned}
\end{equation}
where $\chi(\alpha,\phi)$ is some configuration-space time function [e.g., a solution of~\eref{config-time-function-flrw}] that can also be used as a coordinate in configuration space. The factor of $J$ is the Jacobian determinant $\del{(\chi,F)}{(\alpha,\phi)}$ for the coordinate transformation $(\alpha,\phi)\mapsto (\chi,F)$. Here, $F(\alpha,\phi)$ is a configuration-space function, independent of $\chi(\alpha,\phi)$, such that $(\alpha,\phi)\mapsto (\chi, F)$ is an invertible coordinate transformation. The discrete label $\sigma$ corresponds to the restriction of the physical states such that $\mathrm{sgn}(\hat{p}_{\chi})$ has a definite value. In the case of a relativistic particle, this corresponds to the restriction to the positive- or negative- frequency sector, whereas it corresponds to the restriction to a classically expanding or contracting sector in the de Sitter case [cf.~\eref{pa-de-sitter}] (see~\cite{Chataig:2019-2,Chataig:2020} for a more detailed discussion). In what follows (in particular, Sec.~\ref{sec:expansion}), we will restrict the states to a definite $\sigma$-sector and we will omit the summation sign as well as the $\sigma$ superscript in~\eref{FP-IP-flrw} for brevity. Moreover, the delta function in~\eref{FP-IP-flrw} fixes the gauge in which the time variable is defined by the level sets of $\chi(\alpha,\phi)$ and the operator $\hat{\mu}$ is to be determined by the following criteria. The inner product~\eref{FP-IP-flrw} should: (1) be positive-definite; (2) be conserved with respect to $t$ (unitarity).

In this way, the inner product~\eref{FP-IP-flrw} can be seen as an operator analogue of the Faddeev-Popov gauge-fixing procedure; in particular, $\hat{\mu}$ is analogous to the Faddeev-Popov determinant\footnote{In~\cite{Chataig:2019-2,Chataig:2020} an operator analogous to an invariant extension of the Faddeev-Popov determinant was used. Here, we do not require that $\hat{\mu}$ be an invariant, i.e., to commute with the constraint operator.} (see Sec.~\ref{sec:relation-gf} for further comments on this analogy). We consider that a certain choice of $\chi(\alpha,\phi)$ corresponds to a well-defined gauge in the quantum theory if~\eref{FP-IP-flrw} can be consistently defined from criteria (1) and (2). 

Similar procedures to gauge-fix the inner product were advocated in~\cite{Barvinsky,Woodard:1993,HT:book,Greensite:1989}. In particular, the issue of unitarity and the connection to path integrals were carefully considered in~\cite{Barvinsky} up to one-loop order (i.e., up to order $\hbar$). In the present article, rather than performing an expansion in $\hbar$, we will use the weak-coupling expansion of the BO approach [cf.~Sec.~\ref{sec:expansion}] and we will see how $\hat{\mu}$ can be defined order by order in $\kappa$. The physical Hilbert space of states is, as before, defined as the vector space of superpositions of $\Psi_{E=0,\n}(\alpha,\phi)$ that are square-integrable with respect to~\eref{FP-IP-flrw}.

The advantage of using~\eref{FP-IP-flrw} is that its connection to the dynamics is straightforward: instants of time are defined by the level sets of $\chi(\alpha,\phi)$, provided the gauge is well-defined. Can we offer a \emph{physical} interpretation of what these level sets mean in the quantum theory? In other words, what does $t$ mean in the quantum theory? From~\eref{FP-IP-flrw}, we note that one can define the probabilities
\begin{equation}\label{eq:CP-flrw}
p_{\Psi} :=\frac{1}{\roundprod{\Psi}{\Psi}}\left.\left(\hat{\mu}^{\frac12}\Psi\right)^{*}\hat{\mu}^{\frac12}\Psi\right|_{\chi(\alpha,\phi) = t} \,.
\end{equation}
We suggest that~\eref{CP-flrw} should be interpreted as conditional probabilities, i.e., the probabilities of observing a certain value of $F(\alpha,\phi)$ given the condition that the quantity $\chi(\alpha,\phi)$ is observed (via a measurement) to have the value $t$. A similar remark was made in~\cite{Barvinsky} and the reader is also referred to more recent investigations~\cite{Hoehn:Trinity, Chataig:2020,Hoehn:2020} that discuss the connection between conditional probabilities, relational observables and notions of quantum reference frames. Thus, we take $t$ to be the observed value of a quantity $\chi(\alpha,\phi)$, conditioned on which one makes observations of other quantities.

In principle, one can understand the connection of~\eref{FP-IP-flrw} and its associated notion of dynamics to a notion of relational observables in the following way. Relational observables correspond to linear transformations in the physical Hilbert space (they are on-shell operators) and, thus, they commute with the constraint operator $\hat{C}$. The kinematical operators $\hat{\alpha},\hat{\phi}, \hat{p}_{\alpha}, \hat{p}_{\phi}$, on the other hand, do not commute with $\hat{C}$. Nevertheless, one can define the matrix elements of relational observables by inserting kinematical operators into~\eref{FP-IP-flrw}, i.e., one defines
\begin{equation}\label{eq:relObs-flrw}
\begin{aligned}
&\left(\Psi_{(1)}\left|\hat{\Ob}[f|\chi=t]\right|\Psi_{(2)}\right)\\
&:=\int_{\mathbb{R}^2}\D\alpha\D\phi\ \left(\hat{\mu}^{\frac12}\Psi_{(1)}\right)^{*}\hat{f}|J|\delta(\chi(\alpha,\phi)-t)\hat{\mu}^{\frac12}\Psi_{(2)} \,,
\end{aligned}
\end{equation}
where $\hat{f}$ is a function of the kinematical operators. The relational observable $\hat{\Ob}[f|\chi=t]$ is interpreted as the quantity $\hat{f}$ described with respect to the time variable defined by $\chi(\alpha,\phi)$. A similar definition of relational observables was analysed in~\cite{Chataig:2019-2,Chataig:2020} (see also~\cite{Woodard:1985} for an earlier discussion).

It is worthwhile to note that a definition of the inner product such as~\eref{FP-IP-flrw} carries the assumption that it is, in fact, possible to define a notion (or notions) of time in the quantum theory of reparametrization-invariant systems. This goes against the popular view that the absence of an external or preferred time parameter in~\eref{WDW-flrw} implies that the quantum theory is timeless and that time should emerge only in an appropriate limit (such as the weak-coupling limit used in the BO approach that will be analysed in Sec.~\ref{sec:expansion})~\cite{Kiefer:book,DeWitt:1967}. Even though our usual experience of time is undoubtedly associated with a classical gravitational background (spacetime) and the interpretation of readings of a fully quantum clock is currently controversial or unclear, restricting the notion of time to the (semi)classical level [i.e., to (semi)classical gravitational fields] may be too restrictive. 

Indeed, the definition of the dynamics via~\eref{FP-IP-flrw} and~\eref{relObs-flrw} is well-motivated because: (1) it is analogous to the usual gauge-fixing procedure in quantum Yang-Mills theories, which is justified due to the analogy between a choice of gauge in canonical Yang-Mills theories and a choice of coordinates in canonical general relativity; (2) it extends to the quantum realm one of the key aspects of the classical theory, which is that physically interesting reparametrization-invariant observables are relational quantities, i.e., they represent values of physical fields in relation to the level sets of other dynamical quantities. In the same way, the dynamics defined via~\eref{FP-IP-flrw} and~\eref{relObs-flrw} is relational in the sense that it is conditioned on the value of a time function $\chi(\alpha,\phi)$; (3) the physical meaning of the gauge fixing procedure can be tentatively interpreted in terms of conditional probabilities, following the discussion after~\eref{CP-flrw}.

It is not an issue to use the adjective ``timeless'' to signal the absence of an \emph{external} or \emph{preferred} time parameter. Nevertheless, it may be too restrictive to consider that the quantum theory is \emph{strictly timeless}, i.e., that there is no notion of a repameratrisation-invariant quantum dynamics. With a definition of the inner product such as~\eref{FP-IP-flrw} and of relational observables such as~\eref{relObs-flrw} (see also~\cite{Hoehn:Trinity, Chataig:2019-2, Chataig:2020,Hoehn:2020}), one endeavors to achieve a notion of quantum dynamics that is not strictly timeless, but rather relational, as is the classical dynamics. In other words, in analogy to the classical theory, one assumes that it is in principle possible to parametrize the quantum dynamics with respect to a given choice of time coordinate. 

This view supporting the concept of time beyond the semiclassical level was also expressed in~\cite{Chataig:2019-1} in a different way, where it was also argued that the results of the BO approach coincide with a choice of gauge fixing of the time variable both at classical and quantum levels. A similar argument in favor of extending the concept of time to the purely quantum level as well as a definition of the inner product similar to~\eref{FP-IP-flrw} were given in~\cite{Greensite:1989,Barvinsky}.

\subsection{\label{sec:pert}Classical perturbations}
Let us now consider perturbations to the FLRW background metric. We summarize the main results needed for the analysis of the quantum dynamics in Sec.~\ref{sec:expansion}. The reader is referred to~\cite{BKK-1,BKK-2} and to the usual treatments~\cite{Mukhanov, Halliwell:1984,Langlois:1994,Malk} for further details.

We describe scalar perturbations with four spacetime functions $A$, $B$, $\psi$ and $E$, and tensor perturbations with a symmetric spatial tensor $h_{ij}$. The perturbed line element is
\begin{equation}\label{eq:perturbed-FLRW}
\begin{aligned}
\D s^2 = &a^2(\eta)\curlybr{-(1-2A)\D\eta^2+2(\partial_i B)\D x^i\D\eta\right.\\
&\left.+\squarebr{(1-2\psi)\delta_{ij}+2\partial_i\partial_j E+h_{ij}}\D x^i\D x^j} ,
\end{aligned}
\end{equation}
where we have adopted the conformal time variable $t = \eta$, for which the lapse function $N(\eta) = a(\eta)$. Moreover, we have returned to a dimensionless scale factor and a dimensionful $\eta$ in~\eref{perturbed-FLRW}. The redefinitions~\eref{Lfrak-redef} will be repeated below. In addition to the perturbations of the line element, we also include the scalar perturbations $\varphi(\eta,\x)$ of the inflaton field $\phi(\eta,\x)$.

At the lowest order in the perturbations, it is convenient to work with variables that are invariant under the linearized diffeomorphism symmetry. These variables are usually referred to as ``master gauge-invariant variables''~\cite{BKK-1}. One well-known example is the Mukhanov-Sasaki variable~\cite{Mukhanov,BKK-1}
\begin{equation}\label{eq:ms-variable}
v:=a\curlybr{\varphi+\frac{\dot{\phi}}{\mathcal{H}}\squarebr{A+2\mathcal{H}(B-\dot{E})+\frac{\D}{\D\eta}(B-\dot{E})}} ,
\end{equation}
where $\cdot\equiv\D/\D\eta$ and $\mathcal{H} = \dot{a}/a$. Furthermore, tensor perturbations are already gauge invariant (i.e., invariant under the linearized symmetry) and feature two independent physical degrees of freedom, which are the two polarizations $+,\times$ of gravitational waves.

The dynamics of $v$ and $h_{ij}$ may be derived by expanding the action up to quadratic order in the perturbations,
\begin{equation}\label{eq:action-expansion}
S = S_0+\delta S+\delta^2S+\ldots \,, 
\end{equation}
where $S_0$ is the background action and $\delta S$ vanishes when the background equations of motion are satisfied. Variation of the last term in~\eref{action-expansion} with respect to the perturbative variables yields the equations of motion for the perturbations on a fixed FLRW background. To compute this term, it is useful to consider the Fourier transform of the gauge-invariant perturbations. This is acceptable because all the modes evolve independently at this order of the perturbations. For the Mukhanov-Sasaki variable, one obtains
\begin{equation}\label{eq:ms-fourier}
v(\eta,\x) = \int_{\mathbb{R}^3}\frac{\D^3k}{(2\pi)^{\frac32}}v_{\K}(\eta)\e^{\I\K\cdot\x} \,,
\end{equation}
where we impose the reality condition $v^{*}_{\K} = v_{-\K}$. For the tensor perturbations, we define the rescaled Fourier coefficient for each polarization
\begin{equation}\label{eq:tensor-fourier}
v_{\K}^{(+,\times)} := \frac{a}{\sqrt{12\kappa}}h_{\K}^{(+,\times)} \,.
\end{equation}
With the definitions~\eref{ms-fourier} and~\eref{tensor-fourier}, we find~\cite{BKK-1,Martin-Vennin,Mukhanov}
\begin{equation}\label{eq:action-perts}
\begin{aligned}
\delta^2S &= \int\D\eta\int\D^3k\ \bigg\{\dot{v}_{\K}\dot{v}^{*}_{\K}-\omega_{\K;\mathrm{S}}^2|v_{\K}|^2\\
&+\sum_{\lambda = +,\times}\squarebr{\dot{v}_{\K}^{(\lambda)}\left(\dot{v}_{\K}^{(\lambda)}\right)^{*}-\omega_{\K;\mathrm{T}}^2\left|v_{\K}^{(\lambda)}\right|^2}\bigg\} \,,
\end{aligned}
\end{equation}
where the integral over $\K$ is restricted to half of the Fourier space. In~\eref{action-perts}, we defined the frequencies
\begin{equation}\label{eq:frequencies}
\omega_{\K;\mathrm{S}}^2(\eta):= k^2-\frac{\ddot{z}}{z} \,, \quad \omega_{\K;\mathrm{T}}^2(\eta):=k^2-\frac{\ddot{a}}{a} \,,
\end{equation}
and $k = |\K|$, $z := a\dot{\phi}/\mathcal{H}$. Furthermore, we define
\begin{equation}\label{eq:unified-v}
v_{\K}^{(\rho)} := \left\{\begin{array}{ll}
v_{\K} &\text{ for }\rho = \mathrm{S}\,, \\
v_{\K}^{(+)} &\text{ for }\rho = + \,,\\
v_{\K}^{(\times)} &\text{ for }\rho = \times \,,
\end{array}\right.
\end{equation}
and
\begin{equation}\label{eq:unified-freq}
\omega_{\K;\rho}^2 := \left\{\begin{array}{ll}
\omega_{\K;\mathrm{S}}^2 &\text{ for }\rho = \mathrm{S}\,, \\
\omega_{\K;\mathrm{T}}^2 &\text{ for }\rho = +,\times\,,
\end{array}\right.
\end{equation}
for convenience. The use of~\eref{unified-v} and~\eref{unified-freq} will allow us to work with more compact formulae.

In the quantum theory, it will be advantageous to analyse the dynamics of each Fourier mode separately. To this end, we follow~\cite{BKK-1,BKK-2} and make the replacement
\begin{equation}\label{eq:int-sum-Lfrak}
\int\D^3k \to \frac{1}{\mathfrak{L}^3}\sum_{\K} \ ,
\end{equation}
where $\mathfrak{L}$ is, as before, an arbitrary length. Equation~\eref{int-sum-Lfrak} is justified if one is working with a compact spatial topology. With these assumptions, the wave modes are discretized. Additionally, we make the redefinitions~\eref{Lfrak-redef} and~\cite{BKK-1,BKK-2,Bologna:2014}
\begin{equation}\label{eq:Lfrak-redef-2}
k\mapsto\frac{1}{\mathfrak{L}}k \,, \quad v_{\K}^{(\rho)}\mapsto\mathfrak{L}^2v_{\K}^{(\rho)} \,,
\end{equation}
such that~\eref{action-perts} becomes
\begin{equation}\label{eq:action-perts-2}
\begin{aligned}
\delta^2S &= \int\D\eta\ \sum_{\K,\rho}\ \squarebr{\dot{v}_{\K}^{(\rho)}\left(\dot{v}_{\K}^{(\rho)}\right)^{*}-\omega_{\K;\rho}^2\left|v_{\K}^{(\rho)}\right|^2} ,
\end{aligned}
\end{equation}
where $\rho = \mathrm{S}, +,\times$ and we used~\eref{unified-v} and~\eref{unified-freq}. Finally, it will also be convenient to work with real variables. We thus define~\cite{Martin-Vennin}
\begin{equation}\label{eq:v-R-I}
v_{\K}^{(\rho)} = \frac{1}{\sqrt{2}}\squarebr{v_{\K;\mathrm{R}}^{(\rho)}+\I v_{\K;\mathrm{I}}^{(\rho)}} ,
\end{equation}
where $v_{\K;\mathrm{R}}^{(\rho)}$ and $v_{\K;\mathrm{I}}^{(\rho)}$ are real. From~\eref{action-perts-2}, we then find the Hamiltonian for the perturbations
\begin{equation}\label{eq:H-perts}
H := \frac{1}{2}\sum_{\K,\rho}\sum_{j = \mathrm{R},\mathrm{I}}\curlybr{\squarebr{\pi_{\K;j}^{(\rho)}}^2+\omega_{\K;\rho}^2\squarebr{v_{\K;j}^{(\rho)}}^2}  ,
\end{equation}
where $\pi_{\K;j}^{(\rho)} = \dot{v}_{\K;j}^{(\rho)}$ ($j = \mathrm{R,I}$). It will also be useful to adopt the condensed notation $\q := (\K,j,\rho)$, $v_{\q}:= v_{\K;j}^{(\rho)}$, $\omega_{\q}:=\omega_{\K;\rho}$.

\subsection{\label{sec:master-WDW}Master Wheeler-DeWitt equation}
The quantum theory of perturbations on a fixed, classical FLRW background is obtained by the canonical quantization of~\eref{H-perts} and leads to a (Schr\"{o}dinger-picture) QFT on a curved background, governed by the Schr\"{o}dinger equation
\begin{equation}\label{eq:schrodinger-QFT-curved}
\I\del{\tilde{\psi}}{\eta} = \hat{H}\tilde{\psi} \,,
\end{equation}
where [cf.~\eref{H-perts}]
\begin{align}
\hat{H} &:=\sum_{\q}\hat{H}_{\q} \,,\label{eq:H-perts-quantum}\\
\hat{H}_{\q} &:= \frac12\curlybr{-\del{^2}{v_{\q}^2}+\omega_{\q}^2 v_{\q}^2} .\label{eq:H-perts-quantum-k}
\end{align}
It is at this level that predictions for cosmological observables and the CMB anisotropy spectrum are usually made. However, we are interested in the case in which the background is not classical and is also subject to quantum fluctuations~[cf.~Sec.~\ref{sec:quantum-flrw}]. To deal with this case, one could consider that the wave function must solve the background constraint [in this case,~\eref{WDW-flrw}] in addition to the Schr\"{o}dinger equation~\eref{schrodinger-QFT-curved}.\footnote{As we are keeping terms up to second order in the perturbations, one should in general also impose linearized constraints $\widehat{\delta C}$ on the wave function, which are of first order in the perturbations (while the background constraint~\eref{WDW-flrw} is of zeroth order and the Hamiltonian~\eref{H-perts} is of second order). These linearized constraints can be derived from the $\delta S$ term in~\eref{action-expansion} and were considered in~\cite{Halliwell:1984}, for example. However, the linearized constraints are trivialized for the gauge-invariant perturbations with which we are working. Thus, upon quantization, it is not necessary to impose $\widehat{\delta C}$ on the wave function. This corresponds to a ``reduced phase-space'' quantization of the perturbations (but not of background variables). The reader is referred to~\cite{BKK-1} for further details and references on this issue.} In this way, the quantum dynamics of the background variables and of the perturbations would be dictated by separate equations.

Another possibility is to consider that the background and perturbations form a single reparametrization-invariant system, governed by the master WDW equation
\begin{equation}\label{eq:master-WDW}
\begin{aligned}
\Omega_0\bigg\{\frac{\e^{-3\alpha}}{a_0^3}\bigg[&\frac{\kappa}{2}\del{^2}{\alpha^2}-\frac{1}{2}\del{^2}{\phi^2}\\
&+a_0^6\e^{6\alpha}\mathcal{V}(\phi)\bigg]+\frac{\e^{-\alpha}}{a_0}\hat{H}\bigg\}\Psi(\alpha,\phi,v) = 0 \,,
\end{aligned}
\end{equation}
where the $v$-dependence of the wave function is a short-hand for $v_{\q}$ and $\hat{H}$ is given by~\eref{H-perts-quantum} and~\eref{H-perts-quantum-k}.  The factor of $\e^{-\alpha}/a_0$ that multiplies $\hat{H}$ in~\eref{master-WDW} corresponds to the inverse lapse function for the conformal time variable (with respect to which $\hat{H}$ was constructed) and it is included because the the constraint corresponds to the canonical Hamiltonian divided by the lapse function [cf.~\eref{flrw-action}]. As in~\eref{WDW-flrw}, we also include an overall factor of $\Omega_0$, understood as a choice of einbein frame. This is relevant for the definition of the auxiliary inner product as in~\eref{auxiliary-IP-flrw} and, furthermore, will be of importance in Sec.~\ref{sec:expansion}.

The use of the single master WDW equation~\eref{master-WDW} has the advantage that it directly leads to the usual QFT on a curved, classical background in the weak-coupling limit that will be analyzed in Sec.~\ref{sec:expansion}. Moreover, the weak-coupling expansion also leads to corrections to the Schr\"{o}dinger equation, the unitarity of which will also be assessed in Sec.~\ref{sec:expansion}. In this sense, the use of~\eref{master-WDW} not only incorporates the interaction of the perturbations with a \emph{quantum} background, but it furthermore allows us to go beyond QFT on a curved background in a systematic way, via the weak-coupling expansion.

A master WDW equation has been previously used in~\cite{Halliwell:1984,BKK-1,BKK-2,Bologna:2013,Bologna:2014,Bologna:2016} and it is often analyzed in conjunction with the weak-coupling expansion. What we believe is currently lacking in the literature is a proper understanding of the unitarity of the theory of the master WDW equation beyond the limit of QFT on a classical background [which is obtained in the lowest order of the expansion (cf.~Sec.~\ref{sec:expansion})] and the relation (if any) of this theory to a notion of gauge fixing the time variable (as discussed in Sec.~\ref{sec:quantum-flrw}). These issues are our main focus in this article and we address them below.

We may define an inner product on the space of solutions of the master WDW equation~\eref{master-WDW} in a manner similar to the definition~\eref{FP-IP-flrw}. Given a choice of time function $\chi(\alpha,\phi,v)\equiv \chi$, we define
\begin{equation}\label{eq:FP-IP-flrw-v}
\begin{aligned}
&\roundprod{\Psi_{(1)}}{\Psi_{(2)}}\\
&:= \int\D\alpha\,\D\phi\,\D v\ \left(\hat{\mu}^{\frac12}\Psi_{(1)}\right)^{*}|J|\delta(\chi-t)\hat{\mu}^{\frac12}\Psi_{(2)} \,,
\end{aligned}
\end{equation}
where $J = \del{(\chi,\mathbf{F})}{(\alpha,\phi,v)}$ is the Jacobian for the coordinate transformation $\alpha,\phi,v\mapsto \chi, \mathbf{F}$, and we adopted the short-hand notation
\begin{equation}\label{eq:Dv}
\D v\equiv\prod_{\q}\D v_{\q}\equiv\prod_{\K,\rho,j}\D v_{\K;j}^{(\rho)} \,.
\end{equation}
With the definition~\eref{FP-IP-flrw-v}, we can define the probabilities as in~\eref{CP-flrw}, i.e.,
\begin{equation}\label{eq:CP-flrw-v}
p_{\Psi} :=\frac{1}{\roundprod{\Psi}{\Psi}}\left.\left(\hat{\mu}^{\frac12}\Psi\right)^{*}\hat{\mu}^{\frac12}\Psi\right|_{\chi(\alpha,\phi,v) = t} \,.
\end{equation}
As before, we interpret~\eref{CP-flrw-v} as conditional probabilities, i.e., probabilities of observing certain configurations ${\mathbf{F}}(\alpha,\phi,v)$ of the system $(\alpha,\phi,v)$ given the condition that the quantity $\chi(\alpha,\phi,v)$ is observed to have the value $t$.

Let us now specialize to the de Sitter case. As in the classical theory, we consider the condition $\phi = \phi_0 =$ const., which can be imposed by using probabilities that are conditioned not only on the value of the time function $\chi$, but also on the constant value of the inflaton $\phi = \phi_0$. We thus obtain [cf.~\eref{CP-flrw-v}]
\begin{equation}\label{eq:CP-flrw-v-deSitter}
p_{\Psi} :=\frac{\left.\left(\hat{\mu}^{\frac12}\Psi\right)^{*}\hat{\mu}^{\frac12}\Psi\right|_{\chi = t,\phi = \phi_0}}{\left(\Psi\left|\hat{\Ob}[P_{\phi_0}|\chi=t]\right|\Psi\right)} \,,
\end{equation}
where~[cf.~\eref{relObs-flrw}]
\begin{equation}\label{eq:relObs-Pphi}
\begin{aligned}
&\left(\Psi_{(1)}\left|\hat{\Ob}[P_{\phi_0}|\chi=t]\right|\Psi_{(2)}\right)\\
&:=\int\D\alpha\D\phi\D v\ \left(\hat{\mu}^{\frac12}\Psi_{(1)}\right)^{*}|J|\delta(\phi-\phi_0)\delta(\chi-t)\hat{\mu}^{\frac12}\Psi_{(2)}
\end{aligned}
\end{equation}
corresponds to the insertion of the observable associated with the kinematical improper projector
\begin{equation}
\left<\phi'\left|\hat{P}_{\phi_0}\right|\phi\right>:=\delta(\phi'-\phi)\delta(\phi-\phi_0) \,.
\end{equation}
Moreover, the quantum analogue of the classical condition~\eref{no-roll-flrw} is given by the supplementary condition on the wave function
\begin{equation}\label{eq:no-roll-quantum}
\del{\Psi}{\phi} = 0 \ ,
\end{equation}
whereas the potential $\mathcal{V}(\phi)$ is still given by~\eref{DeSitterCase-V}. In this case, the master WDW equation~\eref{master-WDW} becomes
\begin{equation}\label{eq:master-WDW-deSitter}
\begin{aligned}
\bigg\{\frac{\e^{-2\alpha}}{a_0^2}\bigg[&\frac{\kappa}{2}\del{^2}{\alpha^2}+a_0^6\e^{6\alpha}\frac{H_0^2}{2\kappa}\bigg]+\hat{H}\bigg\}\Psi(\alpha,v) = 0 \,.
\end{aligned}
\end{equation}
A final comment about the use of $\hat{H}$ in~\eref{master-WDW} and~\eref{master-WDW-deSitter} is in order. The frequencies $\omega_{\q}$ are complicated expressions of the background variables and their derivatives  [cf.~\eref{frequencies} and~\eref{unified-freq}] and, therefore, their quantization in a master WDW equation is a delicate matter. In principle, one could define $\omega_{\q}$ as operators that depend on the background variables and their conjugate momenta. This would involve a factor-ordering ambiguity, which, nevertheless, could be a further source of departure from the limit of QFT in a classical spacetime. However, if one defines conformal time as a configuration-space time function [as in~\eref{config-time-function-flrw} and~\eref{conformal-time}], another option is to simply consider that the frequencies are c-number functions of the background variables and of conformal time (also understood as a function of the background variables). This option is available (and more convenient) in the weak-coupling expansion of the BO approach and will be adopted in the next section.\footnote{This simplification was also adopted in~\cite{BKK-0-0,BKK-0-1,BKK-1,BKK-2,BKK-3}.} Indeed, the frequencies in the case of a de Sitter background simplify to [cf.~\eref{frequencies} and~\eref{unified-freq}] 
\begin{equation}\label{eq:freq-deSitter}
\omega^2_{\q}\equiv \omega_{\K}^2 := k^2-\frac{2}{\eta^2} \,,
\end{equation}
where $\eta\equiv\eta(a)$ according to~\eref{conformal-time} (for a fixed choice of $\sigma$). Thus, in the BO approach of the next section, we take $\omega^2_{\q}$ to be a (configuration-space) function of the scale factor.

In what follows, we will solve~\eref{master-WDW-deSitter} using a weak-coupling expansion in powers of $\kappa$ and we will assess the unitarity of the theory. The measure $\hat{\mu}$ that defines the conditional probabilities~\eref{CP-flrw-v-deSitter} will be determined order by order in perturbation theory and its connection to a choice of $\Omega_0$ will be discussed in Sec.~\ref{sec:relation-gf}.

\section{\label{sec:expansion}Weak-coupling expansion and Unitarity}
In the BO approach to quantum gravity and cosmology, one assumes that the degrees of freedom of the time-reparametrization-invariant system can be divided into a set of ``heavy'' or background variables (e.g., the scale factor of the universe and, possibly, the homogeneous mode of the inflaton), associated with an energy scale $M$ (e.g., the Planck mass), and a set of ``light'' variables (e.g., cosmological perturbations), associated with energy scales $m\ll M$. In this case, it is possible to solve the quantum constraint equation using a formal perturbative expansion in powers of $\frac{1}{M}$ if the background potential is non-vanishing. This is what we refer to as the weak-coupling expansion. The reader is referred to~\cite{Kiefer:1993-2,Chataig:2019-1} for reviews and to~\cite{BKK-1,BKK-2,BKK-3,Bologna:2013,Bologna:2014,Bologna:2016} for phenomenological applications.

As we shall see below, a time function is automatically singled out in the BO approach. It corresponds to a choice of time variable for the classical trajectories of the background variables and it governs the evolution of the light degrees of freedom. For this reason, the BO approach is frequently regarded as a solution to the problem of time, in which time emerges in regions of configuration space where the background variables behave semiclassically.\footnote{Nevertheless, it is important to mention that it is possible to consider a certain departure from the purely semiclassical concept of time in the BO approach. The reader is referred to~\cite{Bologna:2018} for earlier work and further details.} This is not, however, the most general solution to the problem of time. Indeed, as we discussed in Secs.~\ref{sec:quantum-flrw} and~\ref{sec:master-WDW}, it may be possible to extend the notion of time and dynamics beyond the semiclassical level in a relational way, by defining the inner product and the dynamics as in~\eref{FP-IP-flrw} and~\eref{FP-IP-flrw-v} (see also the recent articles~\cite{Hoehn:Trinity,Chataig:2019-2,Chataig:2020,Hoehn:2020}). Although this extension of the reparametrization-invariant dynamics to the purely quantum realm is of course still provisional, it may be fruitful to investigate its phenomenological consequences.

On the other hand, even if the BO approach can be seen as a particular case of a more general formalism, it is nevertheless relevant for quantum gravity phenomenology. This is because of the fact that the perturbative expansion in powers of $\frac{1}{M}$ leads to corrections to QFT in curved spacetime. Thus, the BO approach may be of direct interest to the analysis of early-universe phenomena and, in particular, to quantum-gravitational effects in inflationary cosmology~\cite{BKK-1,BKK-2,BKK-3}.
 
We now set out to solve~\eref{master-WDW-deSitter} using the BO approach and we analyse the unitarity of the theory. In Sec.~\ref{sec:corrections}, we apply our results to the calculation of quantum-gravitational corrections to the primordial power spectra of the cosmological perturbations.

\subsection{The minimal BO ansatz}
The system described by the master WDW equation~\eref{master-WDW-deSitter} is precisely of the kind assumed by the BO approach. The heavy variable is $\alpha$, whereas the cosmological perturbations constitute the light degrees of freedom. The background potential is non-vanishing and proportional to~\eref{DeSitterCase-V}. The expansion parameter is the inverse rescaled Planck mass $\kappa = 1/M$. Moreover, following the discussion on the definition of the frequencies $\omega_{\q}$ in Sec.~\ref{sec:master-WDW}, we take the Hamiltonian $\hat{H}$ to depend on $\alpha$ parametrically and to be of order $\kappa^0$.

In this way, to find solutions to~\eref{master-WDW-deSitter}, we make the ``Wentzel-Kramers-Brillouin (WKB)-like'' ansatz
\begin{equation}\label{eq:BO-ansatz}
\Psi(\alpha,v) = \exp\!\left[\frac\I\kappa \, S(\alpha,v)\right] ,
\end{equation}
where $S(\alpha,v)$ is a complex function. We perform a formal expansion of $S(\alpha,v)$ in powers of $\kappa$,
\begin{equation}\label{eq:S-expansion-general}
S(\alpha,v) = \sum_{n = 0}^{\infty}S_n(\alpha,v)\kappa^n =: S_0(\alpha,v) +\kappa\delta S(\alpha,v) \,.
\end{equation}
Likewise, the operator $\hat{\mu}$ in~\eref{CP-flrw-v-deSitter} is expanded as
\begin{equation}\label{eq:BO-measure}
\hat{\mu} \equiv \sum_{n = 0}^{\infty}\kappa^n\hat{\mu}_n\roundbr{\alpha;v,-\I\del{}{v}} ,
\end{equation}
where each coefficient will be defined so as to guarantee that the conditional probabilities depend parametrically on $\alpha$ and are conserved in the time variable to be chosen.

Moreover, it is convenient to \emph{define}
\begin{equation}\label{eq:psim-deltaS-general}
\psi(\alpha;v) := \exp\bigl[\I\delta S(\alpha;v)\bigr]\,, 
\end{equation}
such that~\eref{BO-ansatz} can be written as
\begin{equation}\label{eq:BO-ansatz-1}
\Psi(\alpha,v) = \exp\squarebr{\frac\I\kappa S_0(\alpha,v)}\psi(\alpha;v) \,.
\end{equation}
We refer to~\eref{BO-ansatz-1}, which is equivalent to~\eref{BO-ansatz}, as the minimal BO factorization or ansatz.\footnote{\label{footnote:decoherence}The choice of a single factor of $\e^{\frac\I\kappa S_0(\alpha,v)}$ instead of a superposition such as $\e^{\frac\I\kappa S_0(\alpha,v)}+\e^{-\frac\I\kappa S_0(\alpha,v)}$ can be justified from decoherence arguments~\cite{Kiefer:topology}.} This ansatz, in the form of~\eref{BO-ansatz}, was used in~\cite{BKK-0-0,BKK-0-1,BKK-1,BKK-2,Mariam,BKK-3,Stein1,Stein2,Kiefer:1993-2,Singh:1990,Kiefer:1991}.

The minimal BO factorization is not the traditional BO ansatz that is frequently used in nuclear and molecular physics~\cite{Cederbaum:2008,Abedi:2010,Arce:2012} and, occasionally, in quantum cosmology~\cite{Bologna:2013,Bologna:2014,Bologna:2016,Brout:1987-4,Parentani:1997-2,Bertoni:1996}. In the traditional ansatz, one considers separate factors of a general form for the heavy sector and for the light degrees of freedom, $\Psi(\alpha,v) = \psi_0(\alpha)\psi_{\text{BO}}(\alpha;v)$. In Sec.~\ref{sec:traditional-BO-ansatz}, we will show the equivalence of the minimal and traditional BO ans\"{a}tze, also with regard to the definition of the time variable, unitarity and their relation to the inclusion of the so-called backreaction terms. However, we first analyze the issue of unitarity in the minimal BO factorization for clarity.

\subsection{\label{sec:BO-expansion-general}The background time function and conditional probabilities}
If we insert~\eref{BO-ansatz} or~\eref{BO-ansatz-1} into the left-hand side of~\eref{master-WDW-deSitter}, we obtain a power series in $\kappa$ [cf.~\eref{S-expansion-general}]. As the right-hand side of~\eref{master-WDW-deSitter} vanishes, we set the coefficients of each power of $\kappa$ on left-hand side to be separately equal to zero. Up to the lowest order (order $1/\kappa^2$), we find
\begin{equation}\label{eq:master-WDW-km2}
\frac12\sum_{\q}\roundbr{\del{S_0}{v_{\q}}}^2 = 0 \,.
\end{equation}
This condition is fulfilled by assuming that $S_0(\alpha,v)$ is independent of the perturbations, i.e., $S_0(\alpha,v)\equiv S_0(\alpha)$. Using this condition, we subsequently find the Hamilton-Jacobi equation for the de Sitter background
\begin{equation}\label{eq:HJ-background-general}
-\frac{1}{2}\roundbr{\del{S_0}{\alpha}}^2+\frac{a_0^6}{2}\,\e^{6\alpha}H_0^2 = 0
\end{equation}
at the next order (order $1/\kappa$). Thus, we may take $S_0$ to be a real solution to~\eref{HJ-background-general}, such as
\begin{equation}\label{eq:deSitter-HJ-sol}
S_0(\alpha) = -\frac{\sigma a_0^3H_0}{3}\,\e^{3\alpha}+\text{const.} \,, \quad (\sigma = \pm1) \,.
\end{equation}
We note that $S_0/\kappa$ coincides with the on-shell action~\eref{deSitter-on-shell-action}. We will choose $\sigma = 1$, which corresponds to a classically expanding universe [cf.~\eref{pa-de-sitter}]. As mentioned in Sec.~\ref{sec:quantum-flrw}, the restriction of the quantum theory of $\psi(\alpha,v)$ to the $\sigma = 1$ sector is analogous to the restriction of a quantum relativistic particle to the positive-frequency sector~\cite{Chataig:2019-2}.

From~\eref{deSitter-HJ-sol}, classical trajectories may be defined via the equation [cf.~\eref{classical-flrw-C}]
\begin{equation}\label{eq:eom-background-general}
\frac{\D\alpha}{\D t}:= -\frac{N(\tau)}{a_0^3}\,\e^{-3\alpha}\del{S_0}{\alpha} \equiv -\frac{\kappa N(\tau)}{a_0^3}\,\e^{-3\alpha}p_{\alpha} \,,
\end{equation}
where $N(\tau)$ is an arbitrary but non-vanishing lapse function.

In the following orders (for higher powers of $\kappa$), it will be convenient to describe the quantum dynamics of the perturbations $v$ with respect to the background time variable $t$. This can be done if the values of $t$ correspond to level sets of a certain configuration-space time function $\chi$. We will choose the conformal time function given in~\eref{conformal-time}, which can also be obtained from~\eref{eom-background-general} by choosing $N = a_0\e^{\alpha}$. With this choice, the conditional probabilities~\eref{CP-flrw-v-deSitter} read
\begin{equation}\label{eq:BO-CP}
p_{\Psi}\equiv p_{\Psi}(v|\eta = t) =\left.\frac{\left(\hat{\mu}^{\frac12}\Psi\right)^{*}\hat{\mu}^{\frac12}\Psi}{\int\D v\ \left(\hat{\mu}^{\frac12}\Psi\right)^{*}\hat{\mu}^{\frac12}\Psi}\right|_{\eta(a) = t} \,.
\end{equation}
Using~\eref{BO-measure} and~\eref{BO-ansatz-1}, we see that~\eref{BO-CP} depends only on $\psi(\alpha;v)$, i.e.,
\begin{equation}\label{eq:BO-CP-1}
p_{\Psi}(v|a) =\frac{\left(\hat{\mu}^{\frac12}\psi\right)^{*}\hat{\mu}^{\frac12}\psi}{\int\D v\ \left(\hat{\mu}^{\frac12}\psi\right)^{*}\hat{\mu}^{\frac12}\psi} \,.
\end{equation}
For this reason, we interpret $\hat{\mu}^{\frac12}\psi(\alpha;v)$ as a conditional wave function.

\subsection{\label{sec:qft}Time-dependent Schr\"{o}dinger equation}
The lowest orders of the weak-coupling expansion of~\eref{master-WDW-deSitter} lead to the conditions~\eref{master-WDW-km2} (at order $1/\kappa^2$) and~\eref{HJ-background-general} (at order $1/\kappa$). The next orders (order $\kappa^0$ and higher powers) are governed by the following equation for $\psi(\alpha;v)$ [cf.~\eref{BO-ansatz-1}]
\begin{equation}\label{eq:pre-schrodinger-general}
\begin{aligned}
-\I\,\frac{\e^{-2\alpha}}{a_0^2}\del{S_0}{\alpha}\del{\psi}{\alpha} =\hat{H}\psi&+\roundbr{\frac{\I\e^{-2\alpha}}{2a_0^2}\del{^2S_0}{\alpha^2}}\psi\\
&+\frac{\kappa\e^{-2\alpha}}{2a_0^2}\del{^2\psi}{\alpha^2} \,.
\end{aligned}
\end{equation}
In terms of conformal time [cf.~\eref{conformal-time}], we can rewrite~\eref{pre-schrodinger-general} as
\begin{equation}\label{eq:pre-schrodinger-general-1}
\I\del{\psi}{\eta} = \hat{H}\psi+\frac{3\I}{2\eta}\psi+\frac{\kappa}{2}H_0^2\eta^3\del{}{\eta}\roundbr{\eta\del{\psi}{\eta}} .
\end{equation}
Up to order $\kappa^0$, this reduces to the time-dependent Schr\"{o}dinger equation
\begin{equation}\label{eq:schrodinger-general}
\I\del{\psi}{\eta} = \hat{H}\psi+\frac{3\I}{2\eta}\psi+\Ob(\kappa) \,.
\end{equation}
Rather than interpreting the imaginary term as a source of unitarity violation, we can use it to define the lowest-order measure $\hat{\mu}_0$. Indeed, it is straightforward to verify that~\eref{schrodinger-general} can be rewritten as
\begin{equation}\label{eq:schrodinger-general-1}
\I\del{}{\eta}\roundbr{\frac{1}{H_0}|\eta|^{-\frac32}\psi} = \frac{1}{H_0}|\eta|^{-\frac32}\hat{H}\psi+\Ob(\kappa) \,.
\end{equation}
The factor of $1/H_0$ is included for later convenience. If we define
\begin{equation}\label{eq:lowest-order-mu}
\hat{\mu}_0^{\frac12} := \frac{1}{H_0}|\eta(\alpha)|^{-\frac32} \,,
\end{equation}
then the conditional probabilities~\eref{BO-CP-1} are conserved up to order $\kappa^0$ and reproduce the usual results of QFT on a curved background (see also~Sec.~\ref{sec:corrections}). Indeed, equation~\eref{schrodinger-general-1} is the usual time-dependent Schr\"{o}dinger equation that governs the evolution of the quantum perturbations on a de Sitter background, i.e., it coincides with~\eref{schrodinger-QFT-curved}. The quantum state of the perturbations is given by the conditional wave function $\tilde{\psi}(\alpha;v):=\hat{\mu}_0^{\frac12}\psi(\alpha;v)$, which is to be identified with the wave function used in QFT in curved spacetime, i.e., the solution of~\eref{schrodinger-QFT-curved}.

\subsection{\label{sec:corrected-schrodinger}Corrected Schr\"{o}dinger equation}
We obtain corrections to the (lowest-order) Schr\"{o}dinger equation~\eref{schrodinger-general-1} by using~\eref{pre-schrodinger-general-1} in an iterative fashion. In what follows, we keep terms only up to order $\kappa$. First, we note that~\eref{pre-schrodinger-general-1} is equivalent to
\begin{equation}\label{eq:corrected-schrodinger-0}
\I\del{}{\eta}\roundbr{\frac{|\eta|^{-\frac32}}{H_0}\psi} = \hat{H}\frac{|\eta|^{-\frac32}}{H_0}\psi-\kappa\frac{H_0}{2}|\eta|^{\frac32}\del{}{\eta}\roundbr{\eta\del{\psi}{\eta}} .
\end{equation}
We can rewrite the last term by using~\eref{schrodinger-general}. We obtain
\begin{equation}
\del{}{\eta}\roundbr{\eta\del{\psi}{\eta}} = \del{}{\eta}\roundbr{-\I\eta\hat{H}\psi}-\frac{3\I}{2}\hat{H}\psi+\frac{9}{4\eta}\psi+\Ob(\kappa) \,,
\end{equation}
which leads to
\begin{equation}\label{eq:corrections-0}
|\eta|^{\frac32}\del{}{\eta}\roundbr{\eta\del{\psi}{\eta}} = -\I\del{}{\eta}\roundbr{\eta|\eta|^{\frac32}\hat{H}\psi}+\frac{9|\eta|^{\frac32}}{4\eta}\psi+\Ob(\kappa) \,.
\end{equation}
If we insert~\eref{corrections-0} into~\eref{corrected-schrodinger-0}, we find
\begin{equation}\label{eq:corrected-schrodinger}
\I\del{}{\eta}\roundbr{\hat{\mu}^{\frac12}\psi} = \roundbr{\hat{H}+\kappa\frac{9H_0^2\eta^2}{8}}\frac{|\eta|^{-\frac32}}{H_0}\psi+\Ob(\kappa^2) \,,
\end{equation}
where we defined
\begin{equation}\label{eq:mu-kappa}
\hat{\mu}^{\frac12} := \frac{|\eta|^{-\frac32}}{H_0}\roundbr{1+\frac{\kappa H_0^2\eta^{4}}{2}\hat{H}}+\Ob(\kappa^2) \,.
\end{equation}
It is more convenient to rewrite~\eref{corrected-schrodinger} solely in terms of the conditional wave function $\tilde{\psi} := \hat{\mu}^{\frac12}\psi$. To this end, we use the perturbative inverse of~\eref{mu-kappa},
\begin{equation}\label{eq:mu-kappa-inverse}
\hat{\mu}^{-\frac12} := H_0|\eta|^{\frac32}\roundbr{1-\frac{\kappa H_0^2\eta^{4}}{2}\hat{H}}+\Ob(\kappa^2) \,,
\end{equation}
to obtain from~\eref{corrected-schrodinger} the corrected Schr\"{o}dinger equation
\begin{equation}\label{eq:corrected-schrodinger-eff}
\I\del{\tilde{\psi}}{\eta} = \He\tilde{\psi} \,,
\end{equation}
where
\begin{equation}\label{eq:H-eff-0}
\begin{aligned}
\He &:= \roundbr{\hat{H}+\kappa\frac{9H_0^2\eta^2}{8}}\roundbr{1-\frac{\kappa H_0^2\eta^{4}}{2}\hat{H}}+\Ob(\kappa^2)\\
&=\hat{H}-\kappa\frac{H_0^2\eta^{4}}{2}\hat{H}^2+\kappa\frac{9H_0^2\eta^2}{8}+\Ob(\kappa^2)
\end{aligned}
\end{equation}
is the corrected or effective Hamiltonian for the cosmological perturbations up to order $\kappa$. The last term in~\eref{H-eff-0} only depends on $\eta$ and may be neglected, since it does not affect (conditional) expectation values of the cosmological perturbations and it may be removed by an $\eta$-dependent phase transformation of $\tilde{\psi}$. Thus, we obtain
\begin{equation}\label{eq:H-eff}
\He = \hat{H}-\kappa\frac{H_0^2\eta^{4}}{2}\hat{H}^2+\Ob(\kappa^2) \,.
\end{equation}
Due to the definition of $\hat{H}$ given in~\eref{H-perts-quantum} and~\eref{H-perts-quantum-k}, we see that $\He$ is symmetric with respect to the measure $\D v$ [cf.~\eref{Dv}] and, when suitable boundary conditions are imposed for $\tilde{\psi}$, $\He$ is formally self-adjoint. In this way, we conclude that the evolution of the conditional wave function $\tilde{\psi}$ given in~\eref{corrected-schrodinger-eff} is unitary and the conditional probabilities~\eref{BO-CP-1} are conserved up to order $\kappa$. The phenomenology of~\eref{corrected-schrodinger-eff} will be analysed in Sec.~\ref{sec:corrections}.

Our result for the corrected Hamiltonian~\eref{H-eff} differs from that of~\cite{BKK-1,BKK-2,BKK-3} because no ``unitarity-violating'' terms with imaginary c-number coefficients are present. Such terms have been absorbed into the definition of the measure~\eref{mu-kappa}. The need to define a non-trivial measure was previously pointed out by L\"{a}mmerzahl~\cite{Lammerzahl:1995} in a different context and by one of us in~\cite{Chataig:2019-1}. Here, we have systematically shown how the definition of $\hat{\mu}^{\frac12}$ arises from the weak-coupling expansion. The relation of this measure to the classical Faddeev-Popov determinant will be discussed next. In what follows, we keep terms only up to order $\kappa$ and we omit the $\Ob(\kappa^2)$ remainder for brevity.

\subsection{\label{sec:relation-gf}Relation to a notion of gauge fixing the time variable}
As we discussed in Secs.~\ref{sec:quantum-flrw} and~\ref{sec:master-WDW}, the quantum dynamics of the time-reparametrization-invariant system can be understood via a definition of the inner product and observables [cf.~\eref{FP-IP-flrw},~\eref{relObs-flrw} and~\eref{FP-IP-flrw-v}] that involves a choice of time variable, i.e., the notion of time is defined by the level sets of a certain function $\chi(\alpha,\phi,v)$. In the weak-coupling expansion used in the BO approach, a particular class of time functions is singled out: the background time variables. We have chosen the conformal time~\eref{conformal-time} to parametrize the dynamics. In this case, the definition of probabilities reduces to~\eref{BO-CP-1}.

As the definition of the inner product~\eref{FP-IP-flrw-v} resembles the usual Faddeev-Popov gauge fixing procedure for path integrals, it is worthwhile to analyze how far this analogy goes. In particular, it is interesting to note whether and how the measure $\hat{\mu}$ can be related to the usual Faddeev-Popov determinant defined in~\eref{FP-det}.

Let us examine this issue for the measure defined perturbatively in~\eref{mu-kappa} for the choice of conformal time in the BO approach. The result presented here is, to the best of our knowledge, new. In~\cite{Barvinsky}, Barvinsky presented a construction of a gauge-fixed inner product and its relation to the classical Faddeev-Popov factor using an expansion in $\hbar$, but no connection to the BO approach was established. Here, we use the BO expansion in $\kappa$ rather than $\hbar$.

We begin by noting that the measure $\hat{\mu}_0$ can be related to the configuration space metric. Indeed, using~\eref{conformal-time}, we may rewrite~\eref{lowest-order-mu} as
\begin{equation}\label{eq:mu-0-minimetric-0}
\hat{\mu}_0 = \frac{1}{H_0^2|\eta|^3} = \frac{\sqrt{\kappa}}{a}\left|\del{a}{\eta}\right|\sqrt{|\det\mathcal{G}|} \,,
\end{equation}
where $\det\mathcal{G}$ is the determinant of~\eref{minimetric-a-phi}. Similarly to what was done in~\eref{kin-mu-0} and~\eref{kin-mu} for the auxiliary (kinematical) measure, we perform a conformal transformation [cf.~\eref{mini-conformal}] with the choice of einbein frame $\Omega = a/\sqrt{\kappa}$, such that~\eref{mu-0-minimetric-0} becomes
\begin{equation}\label{eq:mu-0-minimetric}
\hat{\mu}_0 = \left|\del{a}{\eta}\right|\sqrt{|\det\tilde{\mathcal{G}}|} \,,
\end{equation}
which can be interpreted as the square root of the determinant of the (conformally transformed) background minisuperspace metric with respect to the $(\eta,\phi)$ coordinates.

Taking into account the order $\kappa$ terms in~\eref{mu-kappa}, we now rewrite the denominator of~\eref{BO-CP-1} as
\begin{equation}
\int\D v\ \tilde{\psi}_{(1)}^{*}\tilde{\psi}_{(2)} = \int\D v\ \psi^{*}_{(1)}\hat{\mu}\psi_{(2)} \,,
\end{equation}
where we have considered the overlap of two (possibly different) conditional wave functions $\tilde{\psi}_{(1,2)}$ for generality, and we have defined
\begin{equation}\label{eq:mu-kappa2}
\hat{\mu} = \hat{\mu}_0^{\frac12}\roundbr{1+\kappa H_0^2\eta^{4}\hat{H}}\hat{\mu}_0^{\frac12} \,.
\end{equation}
Using~\eref{conformal-time},~\eref{BO-ansatz-1},~\eref{deSitter-HJ-sol} and~\eref{schrodinger-general-1}, we obtain
\begin{equation}\label{eq:mu-FP}
\begin{aligned}
\psi^{*}_{(1)}\hat{\mu}\psi_{(2)}& = \psi^{*}_{(1)}\hat{\mu}_0^{\frac12}\roundbr{1+\I\kappa H_0^2\eta^{4}\del{}{\eta}}\hat{\mu}_0^{\frac12}\psi_{(2)}\\
&= \Psi^{*}_{(1)}\hat{\mu}_0^{\frac12}\roundbr{\I\kappa H_0^2\eta^{4}\del{}{\eta}}\hat{\mu}_0^{\frac12}\Psi_{(2)}\\
&= \Psi^{*}_{(1)}\hat{\mu}_0\roundbr{-\kappa H_0^2\eta^{4}\hat{p}_{\eta}}\Psi_{(2)}\,,
\end{aligned}
\end{equation}
where
\begin{equation}\label{eq:p-eta-quantum}
\hat{p}_{\eta} := -\I\hat{\mu}_0^{-\frac12}\del{}{\eta}\hat{\mu}_0^{\frac12}
\end{equation}
is the operator for the momentum conjugate to $\eta$ with respect to the measure $\hat{\mu}_0$~\cite{DeWitt:1957}, which is related to the background configuration space metric in the $(\eta,\phi)$ coordinates [cf.~\eref{mu-0-minimetric}]. Thus, one sees that the last line of~\eref{mu-FP} yields a quantization of the classical Faddeev-Popov determinant $\tilde{\Delta}_{\eta}$ given in~\eref{FP-det-eta-eta}, i.e., 
\begin{equation}\label{eq:conditional-FP}
\begin{aligned}
\int\D v\ \tilde{\psi}_{(1)}^{*}\tilde{\psi}_{(2)} &= \int\D v\ \psi^{*}_{(1)}\hat{\mu}\psi_{(2)}\\
&\equiv \int\D v\ \Psi^{*}_{(1)}\hat{\mu}_0\hat{\tilde{\Delta}}_{\eta}\Psi_{(2)} \,.
\end{aligned}
\end{equation}
It is in this sense that the measure $\hat{\mu}$, the inner product~\eref{relObs-Pphi} and the conditional probabilities~\eref{BO-CP},~\eref{BO-CP-1} recover the usual Faddeev-Popov determinant and, in this sense, are related to a notion of gauge fixing the time variable. Moreover, because the de Sitter configuration space consists solely of the scale factor and the action~\eref{flrw-action} leads to one constraint, there are no physical degrees of freedom in the de Sitter background; i.e., the background is ``pure gauge''. Once the proper time gauge~\eref{FP-det-eta-eta} is fixed, the physical variables are the relational observables that correspond to the perturbations [cf.~\eref{relObs-flrw}].

We note that the operator $\hat{\mu}_0\hat{\tilde{\Delta}}_{\eta}$ in~\eref{conditional-FP} is in general not symmetric with respect to the auxiliary (kinematical) inner product~\eref{auxiliary-IP-flrw}. Nevertheless, this is not relevant because we are interested in the gauge-fixed inner product for conditional wave functions, which takes the simple, positive-definite form $\int\D v\ \tilde{\psi}_{(1)}^{*}\tilde{\psi}_{(2)}$. The role of the Faddeev-Popov operator $\hat{\mu}_0\hat{\tilde{\Delta}}_{\eta}$ is precisely to connect a solution $\Psi(\alpha,v)$ of the master WDW equation (wave function of the universe) to a conditional wave function $\tilde{\psi}(\alpha;v)$.\footnote{In the BO approach, we restrict ourselves to a single choice of $S_0(\alpha)$ from which the background time variable is defined. This implies that the states $\psi_{(1,2)}$ in~\eref{conditional-FP} are related to the physical states via the relation $\Psi_{(1,2)} = \e^{\I S_0/\kappa}\psi_{(1,2)}$ with a fixed choice of $S_0(\alpha)$. See also footnote~\ref{footnote:decoherence}.} 

Finally, we notice that the left-hand side of~\eref{conditional-FP} is positive-definite and conserved due to the unitarity of the dynamics given in~\eref{corrected-schrodinger-eff}. In this way, the inner product~\eref{conditional-FP} satisfies criteria (1) and (2) established in Sec.~\ref{sec:quantum-flrw}.

\subsection{\label{sec:traditional-BO-ansatz}Traditional BO ansatz}
\subsubsection{Basic equations}
Instead of the minimal BO factorization~\eref{BO-ansatz-1}, let us now consider the traditional BO ansatz, which was applied to the wave function of the universe in~\cite{Bologna:2013,Bologna:2014,Bologna:2016,Brout:1987-4,Bertoni:1996}. The traditional factorization reads
\begin{equation}\label{eq:traditional-BO}
\Psi(\alpha,v) = \psi_0(\alpha)\psi_{\text{BO}}(\alpha;v) \,,
\end{equation}
where $\psi_0(\alpha)$ is the background wave function that encodes the dynamics of the heavy variables (in the present case, the scale factor of the universe), whereas $\psi_{\text{BO}}(\alpha;v)$ is the wave function that dictates the evolution of the light variables (here, the cosmological perturbations). The main idea behind~\eref{traditional-BO} is that the light degrees of freedom should follow adiabatically the semiclassical dynamics of the heavy variables. We will see how this can be understood in connection with the minimal BO factorization and the issue of unitarity.

In the traditional BO approach, the dynamics of both $\psi_0(\alpha)$ and $\psi_{\text{BO}}(\alpha;v)$ is determined from the master WDW equation~\eref{master-WDW-deSitter} in a self-consistent fashion.  One first defines the ``backreaction term''~\cite{Kiefer:2018,Kiefer:1987,Mariam,Chataig:2019-1,Stein2}
\begin{equation}\label{eq:master-WDW-backreaction}
\begin{aligned}
\mathfrak{J}(\alpha) := -\frac{\e^{-2\alpha}}{a_0^2\psi_0}\bigg[&\frac{\kappa}{2}\del{^2\psi_0}{\alpha^2}+a_0^6\e^{6\alpha}\frac{H_0^2}{2\kappa}\psi_0\bigg] .
\end{aligned}
\end{equation}
We note that~\eref{master-WDW-backreaction} coincides with the background WDW equation with an additional potential term $\mathfrak{J}(\alpha)$ that encodes the backreaction of the light variables onto the heavy sector. Backreaction is absent in the particular case $\mathfrak{J}(\alpha) = 0$. We will see that $\mathfrak{J}(\alpha)$ is related to the expectation value of the Hamiltonian of cosmological perturbations. This also justifies referring to $\mathfrak{J}(\alpha)$ as the backreaction term.\footnote{The reader is also referred to~\cite{Halliwell:correlations,Paz} for alternative definitions of backreaction through the analysis of Wigner functions and decoherence.}

If we insert~\eref{traditional-BO} into the master WDW equation~\eref{master-WDW-deSitter} and use~\eref{master-WDW-backreaction}, we obtain the equation for $\psi_{\text{BO}}(\alpha;v)$
\begin{equation}\label{eq:psi-BO-eq}
\kappa\frac{\e^{-2\alpha}}{a_0^2}\del{\log\psi_0}{\alpha}\del{\psi_{\text{BO}}}{\alpha}+(\hat{H}-\mathfrak{J})\psi_{\text{BO}}+\frac{\kappa}{2}\frac{\e^{-2\alpha}}{a_0^2}\del{^2\psi_{\text{BO}}}{\alpha^2} = 0 \,.
\end{equation}
Equations~\eref{traditional-BO} and~\eref{psi-BO-eq} are the basic equations of the traditional BO factorization. The time variable is defined from the phase of $\psi_0(\alpha)$ and it is usually argued that the inclusion of the backreaction term $\mathfrak{J}(\alpha)$ leads to a unitary theory~\cite{Bertoni:1996,Bologna:2017,Kiefer:2018}. We will critique this view and see how this can be understood in relation to the results of the minimal BO factorization.

\subsubsection{Backreaction}
Let us now define the BO average of an operator $\hat{O}_{\text{BO}}$ as
\begin{equation}\label{eq:BO-averages}
\braket{\hat{O}}_{\text{BO}} := \frac{\int\D v\ \psi_{\text{BO}}^{*}\hat{\mu}_{\text{BO}}\hat{O}_{\text{BO}}\psi_{\text{BO}}}{\int\D v\ \psi_{\text{BO}}^{*}\hat{\mu}_{\text{BO}}\psi_{\text{BO}}} \,,
\end{equation}
where the operator $\hat{\mu}_{\text{BO}}^{\dagger} = \hat{\mu}_{\text{BO}}$ is a measure to be determined. In the literature~\cite{Bologna:2013,Bologna:2014,Bologna:2016,Brout:1987-4,Bertoni:1996}, it is usually assumed that $\hat{\mu}_{\text{BO}}$ coincides with the identity, but we will argue that a more general definition is needed. An operator $\hat{O}_{\text{BO}}$ is symmetric with respect to $\hat{\mu}_{\text{BO}}$ if it can be written as $\hat{O}_{\text{BO}} := \hat{\mu}_{\text{BO}}^{-\frac12}\hat{O}\hat{\mu}_{\text{BO}}^{\frac12}$ or $\hat{O}_{\text{BO}} := \hat{\mu}_{\text{BO}}^{-1}\hat{O}$ and $\hat{O}^{\dagger} = \hat{O}$.

The BO average of~\eref{psi-BO-eq} then yields
\begin{equation}\label{eq:backreaction-J}
\begin{aligned}
\mathfrak{J}(\alpha) = \kappa&\frac{\e^{-2\alpha}}{a_0^2}\del{\log\psi_0}{\alpha}\left<\del{}{\alpha}\right>_{\text{BO}}+\braket{\hat{H}}_{\text{BO}}\\
&+\frac{\kappa}{2}\frac{\e^{-2\alpha}}{a_0^2}\left<\del{^2}{\alpha^2}\right>_{\text{BO}} .
\end{aligned}
\end{equation}
Thus, $\mathfrak{J}(\alpha)$ is related to the BO average of the Hamiltonian of the cosmological perturbations. It is in this sense that $\mathfrak{J}(\alpha)$ corresponds to a backreaction term. Moreover, if one inserts~\eref{backreaction-J} back into~\eref{psi-BO-eq}, one can show that the BO averages $\left<\partial/\partial\alpha\right>_{\text{BO}}$ and $\left<\partial^2/\partial\alpha^2\right>_{\text{BO}}$ lead to ``fluctuation terms'', i.e., terms of the kind $\kappa(\hat{O}-\braket{\hat{O}}_{\text{BO}})$~\cite{Bertoni:1996,Chataig:2019-1}. The inclusion of these terms leads to corrections to the adiabatic approximation (the lowest order of the traditional BO approach), in which one neglects such fluctuations. From~\eref{backreaction-J}, one sees that the adiabatic approximation coincides with the lowest order of the weak-coupling expansion, whereas the inclusion of the fluctuation terms associated with $\left<\partial/\partial\alpha\right>_{\text{BO}}$ and $\left<\partial^2/\partial\alpha^2\right>_{\text{BO}}$ comprise terms of order $\kappa$ and higher.

\subsubsection{Equivalence to the minimal BO factorization}
The traditional BO factorization is clearly ambiguous. One may perform the redefinitions~\cite{Kiefer:2018}
\begin{equation}\label{eq:Berry-transf}
\begin{aligned}
\psi_0(\alpha) &\mapsto \psi_0(\alpha)\,\e^{\gamma(\alpha)+\I\beta(\alpha)} \ ,\\
\psi_{\text{BO}}(\alpha;v) &\mapsto \psi_{\text{BO}}(\alpha;v)\,\e^{-\gamma(\alpha)-\I\beta(\alpha)} \,,
\end{aligned}
\end{equation}
without altering the wave function of the universe $\Psi(\alpha,v)$. In~\eref{Berry-transf}, $\gamma(\alpha)$ and $\beta(\alpha)$ are real functions\footnote{In particular, the arbitrary phase $\beta(\alpha)$ is related to the Berry phase ambiguity~\cite{Chataig:2019-1}.} that can be expanded in powers of $\kappa$, starting at order $\kappa^0$. Thus, we may write
\begin{equation}\label{eq:traditional-minimal}
\begin{aligned}
\psi_0(\alpha) &= \exp\left[\frac\I\kappa S_0(\alpha)\right]\e^{\gamma(\alpha)+\I\beta(\alpha)} \,, \\
\psi_{\text{BO}}(\alpha;v) &= \psi(\alpha;v)\,\e^{-\gamma(\alpha)-\I\beta(\alpha)} \,,
\end{aligned}
\end{equation}
without loss of generality. If we insert~\eref{traditional-minimal} into~\eref{traditional-BO}, we recover the minimal BO factorization~\eref{BO-ansatz-1}. Moreover, it is straightforward to verify that~\eref{psi-BO-eq} is equivalent to~\eref{pre-schrodinger-general} if one inserts~\eref{traditional-minimal} into~\eref{psi-BO-eq} and eliminates $\mathfrak{J}$ using~\eref{master-WDW-backreaction}. Thus, the results of both factorizations are equivalent. Although this is expected, the ensuing interesting questions are: how does the traditional BO ansatz lead to a unitary evolution? How is this equivalent to the conclusions of Sec.~\ref{sec:corrected-schrodinger}? As the question of the unitarity of the BO approach has been a controversial topic in the literature~\cite{BKK-1,BKK-2,BKK-3,Kiefer:1991}, we believe it is worthwhile to analyze the answers to these questions in the formalism presented here.

\subsubsection{\label{sec:ambiguity-backreaction}Ambiguity of backreaction and unitarity}
As the minimal BO ansatz does not explicitly refer to backreaction, and due to the equivalence between the minimal and traditional BO factorizations, we conclude that unitarity cannot be a consequence of the introduction of the backreaction term alone. Due to the ambiguity~\eref{Berry-transf} in the definition of $\psi_{\text{BO}}(\alpha;v)$, there is also an ambiguity in the definition of the BO averages $\left<\partial/\partial\alpha\right>_{\text{BO}}$, $\left<\partial^2/\partial\alpha^2\right>_{\text{BO}}$ and, consequently, the backreaction term $\mathfrak{J}(\alpha)$ is ambiguous. Indeed, at this stage, $\mathfrak{J}(\alpha)$ is arbitrary. A fixation of $\mathfrak{J}(\alpha)$ corresponds to a fixation of $\gamma(\alpha)$ and $\beta(\alpha)$ in~\eref{traditional-minimal}.

In this way, how can one obtain a unitary evolution for $\psi_{\text{BO}}(\alpha;v)$? Given a pair $\psi_{\text{BO}(1)}$, $\psi_{\text{BO}(2)}$ of solutions to~\eref{psi-BO-eq}, the evolution is unitary if the following condition holds,
\begin{equation}\label{eq:BO-factorization-unitarity}
\del{}{\eta}\int\D v\ \psi_{\text{BO}(1)}^{*}(\eta;v)\hat{\mu}_{BO}\psi_{\text{BO}(2)}(\eta;v) = 0 \,.
\end{equation}
This is equivalent to assuming that~\eref{psi-BO-eq} can be rewritten as
\begin{equation}\label{eq:eff-psi-BO-eq}
\I\del{}{\eta}\left(\hat{\mu}_{\text{BO}}^{\frac12}\psi_{\text{BO}}\right) = \hat{H}_{\text{BO}}\hat{\mu}_{\text{BO}}^{\frac12}\psi_{\text{BO}} \,, 
\end{equation}
for some choice of $\hat{\mu}_{\text{BO}}$ and an operator $\hat{H}_{\text{BO}}$ that is Hermitian with respect to the flat measure $\hat{1}$. Note that this was exactly the procedure shown in Sec.~\ref{sec:corrected-schrodinger} for the minimal BO ansatz, where~\eref{corrected-schrodinger-eff} was the equivalent of~\eref{eff-psi-BO-eq}. This is not, however, the standard procedure considered for the traditional BO ansatz.

In the literature~\cite{Brout:1987-4,Bertoni:1996,Bologna:2017}, the standard procedure focuses on a flat measure $\hat{\mu}_{\text{BO}}\to\hat{1}$ and considers~\eref{BO-factorization-unitarity} for a single state. In this case, the equation~\eref{BO-factorization-unitarity} becomes
\begin{equation}\label{eq:BO-factorization-unitarity-flat}
\begin{aligned}
0 &= \mathfrak{Re}\left.\left<\del{}{\eta}\right>_{\text{BO}}\right|_{\hat{\mu}_{\text{BO}}\to\hat{1}}\\
& = H_0 a_0 \e^{\alpha} \mathfrak{Re}\left.\left<\del{}{\alpha}\right>_{\text{BO}}\right|_{\hat{\mu}_{\text{BO}}\to\hat{1}} .
\end{aligned}
\end{equation}
The task is then to prove that~\eref{BO-factorization-unitarity-flat} is satisfied for solutions of~\eref{psi-BO-eq}. In Sec.~\ref{sec:corrections-unitarity-example}, we give a concrete example of this kind of calculation. It is straightforward to show that (see~\cite{Chataig:2019-1} for details)
\begin{equation}
\mathfrak{Re}\left.\left<\del{}{\alpha}\right>_{\text{BO}}\right|_{\hat{\mu}_{\text{BO}}\to\hat{1}} = \frac12 \del{}{\alpha}\log\left(\int\D v\ \psi_{\text{BO}}^{*}\psi_{\text{BO}}\right).
\end{equation}
Hence~\eref{BO-factorization-unitarity-flat} is satisfied if and only if the norm of $\psi_{\text{BO}}$ is independent of $\alpha$ (unitarity). However, in the standard procedure~\cite{Brout:1987-4,Bertoni:1996,Bologna:2017},  one (sometimes tacitly) makes the assumption that the BO averages
\begin{equation}\label{eq:Berry-connection}
\left.\left<\del{}{\alpha}\right>_{\text{BO}}\right|_{\hat{\mu}_{\text{BO}}\to\hat{1}} = \I A(\alpha) 
\end{equation}
are purely imaginary, where $A(\alpha)$ is the Berry connection related to the Berry phase~\cite{Chataig:2019-1}. But this is precisely what one would need to show. Assuming that~\eref{Berry-connection} is purely imaginary corresponds to assuming that~\eref{BO-factorization-unitarity-flat} is satisfied. Thus, the proof that~\eref{psi-BO-eq} leads to a unitary dynamics becomes circular.

It is worthwhile to mention that in the usual treatment of the BO approximation to nuclear and molecular physics~\cite{BO:1927,Cederbaum:2008,Abedi:2010,Arce:2012}, this problem does not arise because an external time parameter is available in these applications. Even if one is dealing with a stationary Schr\"{o}dinger equation, one is not usually interested in adopting a relational point of view, in which one of the degrees of freedom serves as an internal or intrinsic clock. In contrast, the formalism presented here is relational, and it is our task to determine whether the dynamics is unitary with respect to the chosen time variable.

It is also important to notice that, instead of trying to prove that the dynamics described by~\eref{psi-BO-eq} is unitary, one could simply enforce the condition~\eref{BO-factorization-unitarity-flat} by a suitable choice of $\gamma(\alpha)$ in~\eref{traditional-minimal}. Indeed, one can choose $\gamma(\alpha)$ in such a way that the following relation holds
\begin{equation}\label{eq:enforcing-unitarity-gamma}
\psi_{\text{BO}}(\alpha;v) = \frac{\psi(\alpha;v)\e^{-\I\beta(\alpha)}}{\sqrt{\int\D v\ \psi^*\psi}} \,,
\end{equation}
which guarantees that $\psi_{\text{BO}}(\alpha;v)$ is normalized at all times with respect to the flat measure. This is effectively what is done when one assumes from the start that~\eref{Berry-connection} is purely imaginary. In this case, it is a specific choice of \emph{factorization} of $\Psi(\alpha,v)$ that guarantees the unitarity of the evolution of $\psi_{\text{BO}}(\alpha;v)$. This particular factorization, in turn, corresponds to a (partial) fixation of the backreaction term $\mathfrak{J}(\alpha)$. A similar remark was made by one of us in~\cite{Chataig:2019-1}.

In~\cite{Kiefer:2018}, it was argued that unitarity-violating terms could be neglected because they could be absorbed into redefinitions of the minimal BO factorization according to~\eref{traditional-minimal}. However, as $\psi_0(\alpha)$ is independent of $v$, one cannot absorb $v$-dependent terms present in~\eref{mu-kappa} into a redefinition of $S_0(\alpha)$ in~\eref{traditional-minimal}. Nevertheless, in the formalism of~\cite{Kiefer:2018}, one enforces unitarity at each order of the weak-coupling expansion by a suitable choice of $\gamma(\alpha)$ and, thus, this procedure should correspond\footnote{More precisely, the authors of~\cite{Kiefer:2018} seem to require that $\psi_{\text{BO}}$ is an eigenstate of $\hat{H}$ and that $\I\partial\psi_{\text{BO}}/\partial\eta$ is independent of $v$. We do note make such requirements here.} to the choice~\eref{enforcing-unitarity-gamma}. This option is, however, not preferred because it is non-linearly dependent on the state $\psi(\alpha;v)$. As was remarked in Sec.~\ref{sec:corrected-schrodinger}, a more general procedure is to the absorb the unitarity-violating terms into the measure $\hat{\mu}$ defined in~\eref{mu-kappa}.

Due to the results of Secs.~\ref{sec:corrected-schrodinger},~\ref{sec:relation-gf} and the equivalence of the minimal and traditional BO ans\"{a}tze, we have already established that the BO weak-coupling expansion leads to a unitary evolution of conditional wave functions. In what follows, we give the connection between the measure $\hat{\mu}_{\text{BO}}$ to be used in the traditional BO approach and the measure $\hat{\mu}$ given in~\eref{mu-kappa2}.

\subsubsection{\label{sec:traditional-relation-gf}Relation to the gauge-fixed measure and conditional probabilities}
As~\eref{psi-BO-eq} is equivalent to~\eref{pre-schrodinger-general}, we can establish the unitarity of the dynamics of~\eref{psi-BO-eq}, i.e., the validity of~\eref{BO-factorization-unitarity}, by choosing $\hat{\mu}_{\text{BO}}$ adequately. For a given choice of $\gamma(\alpha)$, let us define
\begin{equation}\label{eq:BO-measure-FP}
\begin{aligned}
\hat{\mu}_{\text{BO}} &\equiv \hat{\mu}_{\text{BO}}\left(\alpha;v,-\I\del{}{v}\right) \\
&:= \e^{2\gamma(\alpha)}\hat{\mu}\left(\alpha;v,-\I\del{}{v}\right) , 
\end{aligned}
\end{equation}
where $\hat{\mu}$ was given in~\eref{mu-kappa2}. Using~\eref{traditional-minimal}, we thus obtain
\begin{equation}
\begin{aligned}
\int\D v\ \psi_{\text{BO}(1)}^{*}\hat{\mu}_{BO}\psi_{\text{BO}(2)} = \int\D v\ \psi_{(1)}^{*}\hat{\mu}\psi_{(2)} \,,
\end{aligned}
\end{equation}
which is equivalent to the conserved~\eref{conditional-FP}. Thus, we find that the overlap with respect to $\hat{\mu}_{\text{BO}}$ of solutions to~\eref{psi-BO-eq} is conserved in conformal time and the dynamics is unitary, provided the weak-coupling expansion is valid and appropriate boundary conditions are imposed for the conditional wave functions. Moreover, the BO averages of operators of the form $\hat{O}_{\text{BO}} := \hat{\mu}_{\text{BO}}^{-\frac12}\hat{O}(\alpha;v,-\I\partial/\partial v)\hat{\mu}_{\text{BO}}^{\frac12}$ are then equivalent to conditional expectation values,
\begin{equation}
\begin{aligned}
\braket{\hat{O}}_{\text{BO}} &= \frac{\int\D v\ \psi_{\text{BO}}^{*}\hat{\mu}_{BO}\hat{O}_{\text{BO}}\psi_{\text{BO}}}{\int\D v\ \psi_{\text{BO}}^{*}\hat{\mu}_{BO}\psi_{\text{BO}}}\\
&= \frac{\int\D v\ \tilde{\psi}^{*}\hat{O}\tilde{\psi}}{\int\D v\ \tilde{\psi}^{*}\tilde{\psi}} \equiv \braket{\hat{O}} .
\end{aligned}
\end{equation}
Finally, if we choose $\beta(\alpha) = 0$ in~\eref{traditional-minimal}, which corresponds to a choice of Berry phase, we obtain $\hat{\mu}_{\text{BO}}^{\frac12}\psi_{\text{BO}} \equiv \tilde{\psi}$. Thus, the solutions to~\eref{psi-BO-eq} are also related to conditional wave functions. This is in line with the formalism proposed by Hunter long ago~\cite{Hunter}, although the main difference is the presence of the non-trivial measure $\hat{\mu}_{\text{BO}}^{\frac12}$ related to the insertion of the Faddeev-Popov operator in the inner product for solutions of the constraint equation [cf.~\eref{conditional-FP}].

To summarize, we have argued that the traditional BO ansatz yields equations that are equivalent to the ones obtained in the minimal BO factorization, also concerning unitarity, provided an adequate choice of measure is made [cf.~\eref{BO-measure-FP}]. We now apply the formalism presented here to the calculation of primordial power spectra.

\section{\label{sec:corrections}Corrections to primordial power spectra}
Up to order $\kappa$, the dynamics of cosmological perturbations is governed by the corrected Schr\"{o}dinger equation~\eref{corrected-schrodinger-eff}, which defines a QFT on a de Sitter background with the effective Hamiltonian~\eref{H-eff}. Let us now compute the corresponding power spectra. At order $\kappa^0$, this will coincide with the usual primordial power spectra in a de Sitter spacetime, whereas terms of order $\kappa$ will yield corrections that originate from the weak-coupling expansion of the master WDW equation~\eref{master-WDW-deSitter}.

\subsection{\label{sec:single-mode}Single-mode equations}
To compute the power spectra, we adopt a simplification that is common in the literature~\cite{BKK-0-0,BKK-0-1,BKK-1,BKK-2,BKK-3,Bologna:2017,Kiefer:1987,Stein2,Mariam}. Namely, we restrict the corrected Schr\"{o}dinger equation~\eref{corrected-schrodinger-eff} to a single Fourier mode. This is sometimes referred to as a ``random phase approximation''~\cite{Kiefer:1987,Stein2,Mariam}.

More precisely, we make the separation ansatz
\begin{equation}\label{eq:k-separation}
\tilde{\psi}(\alpha;v) = \prod_{\q}\tilde{\psi}_{\q}\left(\alpha;v_{\q}\right) ,
\end{equation}
and, using~\eref{H-perts-quantum}, we write
\begin{equation}\label{eq:rpa}
\begin{aligned}
\hat{H}^2\tilde{\psi} &= \sum_{\q}\hat{H}_{\q}^2\tilde{\psi}+\sum_{\q'\neq\q}\hat{H}_{\q}\hat{H}_{\q'}\tilde{\psi}\\
&= \sum_{\q}\hat{H}_{\q}^2\tilde{\psi}+\tilde{\psi}\sum_{\q'\neq\q}\frac{\hat{H}_{\q}\tilde{\psi}_{\q}}{\tilde{\psi}_{\q}}\frac{\hat{H}_{\q'}\tilde{\psi}_{\q'}}{\tilde{\psi}_{\q'}}\\
&= \sum_{\q}\hat{H}_{\q}^2\tilde{\psi}-\tilde{\psi}\sum_{\q'\neq\q}\frac{\partial_{\eta}\tilde{\psi}_{\q}}{\tilde{\psi}_{\q}}\frac{\partial_{\eta}\tilde{\psi}_{\q'}}{\tilde{\psi}_{\q'}}+\Ob(\kappa) \,.
\end{aligned}
\end{equation}
The random phase approximation consists in the assumption that the terms $\partial_{\eta}\tilde{\psi}_{\K}/\tilde{\psi}_{\K}$ add incoherently in such a way that the second term on the right-hand side of~\eref{rpa} can be neglected. It is important to mention that this approximation is a formal procedure. The second term on the right-hand side of~\eref{rpa} will in general exhibit field-theoretic divergences that need to be removed with a subtraction scheme~\cite{Halliwell:1984,Mazzitelli,Eboli}.\footnote{The reader is referred to~\cite{Mazzitelli} (see, in particular, Sec.~IV of~\cite{Mazzitelli}), where the adiabatic subtraction scheme was applied to the WDW equation in the presence of higher-derivative terms in the Born-Oppenheimer approach. The reader is also referred to~\cite{Eboli} for a discussion of the adiabatic subtraction scheme in the functional Schr\"{o}dinger picture for quantum fields on FLRW backgrounds.} To the best of our knowledge, a thorough treatment of the random phase approximation for the master WDW equation~\eref{master-WDW-deSitter} is lacking in the literature. We hope to address the details of this procedure in future work.

What is the physical content of this random phase approximation? It corresponds to assuming that interactions (contained in the $\hat{H}^2$-term) between the different $v_{\q}$ modes are negligible or that a single mode is present in the classical theory and subsequently quantized. As was remarked in~\cite{Brizuela:moment}, by neglecting the interactions between the different $v_{\q}$ modes, we are focusing on the principal difference between the physics of the master WDW equation~\eref{master-WDW-deSitter} and the usual treatment of QFT on curved backgrounds, which is the fact that the de Sitter background is now also quantized.

Note that the neglected interactions include terms of quartic order in the cosmological perturbations and, therefore, a consistent treatment of these terms would require the inclusion of higher-order $\Ob(v^3)$-terms in the classical theory. As the master WDW equation~\eref{master-WDW-deSitter} was based on a perturbative analysis of the classical theory, in which the action was expanded up to quadratic order in the cosmological perturbations, we consider that~\eref{master-WDW-deSitter} and the associated corrected Schr\"{o}dinger equation~\eref{corrected-schrodinger-eff} are only reliable in regions of configuration space where the $\Ob(v^3)$-terms are small and where the random phase approximation is also justified. In what follows, we assume that the formal random phase approximation may be used, i.e., that it is possible to adopt a certain regularization scheme and that the necessary subtractions have been made.

With the random phase approximation applied to~\eref{rpa}, we obtain from~\eref{corrected-schrodinger-eff} the single-mode corrected Schr\"{o}dinger equation
\begin{equation}\label{eq:corrected-schrodinger-single-mode}
\I\del{\tilde{\psi}_{\q}}{\eta} = \hat{H}_{\q}\tilde{\psi}_{\q}-\kappa\frac{H_0^2\eta^{4}}{2}\hat{H}^2_{\q}\tilde{\psi}_{\q} \,,
\end{equation}
where $\hat{H}_{\q}$ was defined in~\eref{H-perts-quantum-k}. 

\subsection{Choice of state}
The precise form of the corrections to the primordial power spectra depends not only on the validity of the assumptions (weak-coupling expansion, random phase approximation) made in the derivation of~\eref{corrected-schrodinger-single-mode}, but also on the choice of initial state. In the present article, we take the position that the state of a reparametrization-invariant system should be defined in a relational way. Following our discussion in Secs.~\ref{sec:master-WDW} and~\ref{sec:BO-expansion-general}, this may be achieved with the use of the conditional wave functions $\tilde{\psi} = \hat{\mu}^{\frac12}\psi$, which yield predictions for the cosmological perturbations conditioned on the value of the scale factor [or of conformal time, understood as function of $a$, cf.~\eref{conformal-time}]. Due to~\eref{BO-ansatz-1} and~\eref{k-separation}, a choice of conditional states $\tilde{\psi}_{\q}$ corresponds to a choice of wave function of the universe, via the formula
\begin{equation}
\Psi(\alpha,v) = \e^{\frac\I\kappa S_0(\alpha)}\hat{\mu}^{-\frac12}\prod_{\q}\tilde{\psi}_{\q}\left(\alpha;v_{\q}\right) .
\end{equation}
We choose the conditional states such that they correspond to the usual Bunch-Davies vacuum at order $\kappa^0$. It is also possible to consider more general choices (such as the excited states considered in~\cite{BKK-3}), but this will not be done here. We thus make the ansatz
\begin{equation}\label{eq:master-ansatz}
\tilde{\psi}_{\q} = \Ncal_{\q}(\alpha)\exp\curlybr{-\frac12\,\Omega_{\q}(\alpha)v_{\q}^2-\frac{\kappa}{4}\,\Gamma_{\q}(\alpha)v_{\q}^4} ,
\end{equation}
where $\Ncal_{\q}(\alpha)$ is a normalization factor and $\mathfrak{Re}\Omega_{\q}(\alpha),\,\mathfrak{Re}\Gamma_{\q}(\alpha) > 0$. We have included the quartic term in~\eref{master-ansatz} for completeness because~\eref{corrected-schrodinger-single-mode} contains terms of quartic order in $v_{\q}$. Moreover, we will see that the function $\Gamma_{\q}(\alpha)$ contributes to the corrections for the power spectra.

Nevertheless, as we remarked earlier, we consider that~\eref{corrected-schrodinger-single-mode} is only reliable in regions where $\Ob(v^3)$-terms are small. Thus, the inclusion of quartic terms need not entail observable non-Gaussianities in the CMB, since~\eref{master-WDW-deSitter} may be significantly modified in regions where such non-Gaussian terms are sizeable. A similar observation was made in the semiclassical treatment of~\cite{deAlwis}.

If we insert~\eref{master-ansatz} into~\eref{corrected-schrodinger-single-mode} and keep terms only up to order $\kappa$, we obtain the equations
\begin{align}
&\I\del{}{\eta}\log\Ncal_{\q} = \frac{\Omega_{\q}}{2}+\frac{\kappa H_0^2\eta^4}{4}\omega_{\q}^2-\frac{3\kappa H_0^2\eta^4}{8}\Omega_{\q}^2 \,, \label{eq:N}\\
&\I\del{\Omega_{\q}}{\eta} = \Omega_{\q}^2-\omega_{\q}^2-3\kappa\Gamma_{\q}-\frac{3\kappa H_0^2\eta^4}{2}\Omega_{\q}\left(\Omega_{\q}^2-\omega_{\q}^2\right) , \label{eq:Om}\\
&\I\del{\Gamma_{\q}}{\eta} = 4\Omega_{\q}\Gamma_{\q}+\frac{H_0^2\eta^4}{2}(\Omega_{\q}^2-\omega_{\q}^2)^2 \,,\label{eq:Ga}
\end{align}
which we will solve in order to ascertain the unitarity of the theory and to compute the corrections to the power spectra.

\begin{widetext}
\subsection{\label{sec:corrections-unitarity-example}Unitarity}
According to the discussion in Sec.~\ref{sec:corrected-schrodinger}, the dynamics described by~\eref{corrected-schrodinger-single-mode} is unitary. We can explicitly verify this for the ansatz~\eref{master-ansatz} as follows. As in Sec.~\ref{sec:ambiguity-backreaction}, we note that the norm of~\eref{master-ansatz} is conserved if
\begin{equation}
0 = \frac{\I}{2}\del{}{\eta}\log\int_{-\infty}^{\infty}\D v_{\q}\ |\tilde{\psi}_{\q}|^2 = \I\,\mathfrak{Im}\left<\I\del{}{\eta}\right> .
\end{equation}
Moreover, using~\eref{master-ansatz}, we obtain the identity
\begin{equation}\label{eq:3-contributions-unitarity-0}
\left<\I\del{}{\eta}\right> = \left<\I\del{}{\eta}\log\Ncal_{\q}-\frac{\I}{2}\del{\Omega_{\q}}{\eta}v_{\q}^2-\frac{\I\kappa}{4}\del{\Gamma_{\q}}{\eta}v_{\q}^4\right> .
\end{equation}
From~\eref{master-ansatz},~\eref{N},~\eref{Om} and~\eref{Ga}, we find\footnote{All integrals are Gaussian because we assume that the quartic term in~\eref{master-ansatz} can be treated perturbatively.}
\begin{equation}\label{eq:3-contributions-unitarity}
\begin{aligned}
\mathfrak{Im}\left<\I\del{}{\eta}\log\Ncal_{\q}\right> &= \frac{\mathfrak{Im}\Omega_{\q}}{2}-\frac{3\kappa H_0^2\eta^4}{4}(\mathfrak{Re}\Omega_{\q})\mathfrak{Im}\Omega_{\q} \,, \\
\mathfrak{Im}\left<-\frac{\I}{2}\del{\Omega_{\q}}{\eta}v_{\q}^2\right> &= -\frac{\mathfrak{Im}\Omega_{\q}}{2}+\frac{9\kappa H_0^2\eta^4}{8}(\mathfrak{Re}\Omega_{\q})\mathfrak{Im}\Omega_{\q}-\frac{3\kappa H_0^2\eta^4 \mathfrak{Im}\Omega_{\q} \left(\omega_{\q} ^2+\mathfrak{Im}\Omega_{\q}^2\right)}{8 \mathfrak{Re}\Omega_{\q}}+\frac{3\kappa (\mathfrak{Re}\Gamma_{\q}) \mathfrak{Im}\Omega_{\q}}{4 \mathfrak{Re}\Omega_{\q}^2}+\frac{3\kappa \mathfrak{Im}\Gamma_{\q}}{4 \mathfrak{Re}\Omega_{\q}} \,,\\
\mathfrak{Im}\left<-\frac{\I}{4}\del{\Gamma_{\q}}{\eta}v_{\q}^4\right> &=-\frac{3 \kappa H_0^2\eta ^4}{8}(\mathfrak{Re}\Omega_{\q})\mathfrak{Im}\Omega_{\q}+\frac{3 \kappa H_0^2\eta ^4\mathfrak{Im}\Omega_{\q}\left(\omega_{\q}^2+\mathfrak{Im}\Omega_{\q}^2\right)}{8 \mathfrak{Re}\Omega_{\q}} -\frac{3 \kappa (\mathfrak{Re}\Gamma_{\q}) \mathfrak{Im}\Omega_{\q}}{4 \mathfrak{Re}\Omega_{\q}^2}-\frac{3 \kappa \mathfrak{Im}\Gamma_{\q}}{4 \mathfrak{Re}\Omega_{\q}} \,,
\end{aligned}
\end{equation}
where we kept terms only up to order $\kappa$. If we add all terms from~\eref{3-contributions-unitarity} and we use~\eref{3-contributions-unitarity-0}, we obtain the expected result
\begin{equation}\label{eq:norm-is-conserved}
\mathfrak{Im}\left<\I\del{}{\eta}\right> = 0 \,.
\end{equation}
\end{widetext}

\subsection{Power spectra}
We define the power spectrum associated with the cosmological perturbations as
\begin{equation}
\mathcal{P}_v(\q) := \frac{k^3}{2\pi^2}\braket{v_{\q}^2} ,
\end{equation}
where we recall that $\q = (\K,j,\rho)$ and
\begin{equation}\label{eq:conditional-correlation-0}
\braket{v_{\q}^2} := \mathrm{E}_{\Psi}[v_{\q}^2|\eta,\phi] = \frac{\left(\Psi\left|\hat{\Ob}[v_{\q}^2P_{\phi}|\eta]\right|\Psi\right)}{\left(\Psi\left|\hat{\Ob}[P_{\phi}|\eta]\right|\Psi\right)}
\end{equation}
is the conditional correlation function for the $\q$-modes. Due to~\eref{FP-IP-flrw-v} and~\eref{relObs-Pphi}, equation~\eref{conditional-correlation-0} can be simplified to the familiar formula
\begin{equation}\label{eq:conditional-correlation}
\braket{v_{\q}^2} = \frac{\int\D v\ \tilde{\psi}^{*}v_{\q}^2\tilde{\psi}}{\int\D v\ \tilde{\psi}^{*}\tilde{\psi}} \,.
\end{equation}
Using~\eref{k-separation} and~\eref{master-ansatz}, we find
\begin{equation}
\braket{v_{\q}^2} = \frac{1}{2\mathfrak{Re}\Omega_{\q}}-\frac{3\kappa\mathfrak{Re}\Gamma_{\q}}{4(\mathfrak{Re}\Omega_{\q})^3} \,.
\end{equation}
Up to order $\kappa$, we may write $\Omega_{\q} = \Omega_{\q;0}+\kappa\Omega_{\q;1}$ to find
\begin{equation}\label{eq:correction-variance-0}
\braket{v_{\q}^2} = \frac{1+\kappa\delta_{\q}}{2\mathfrak{Re}\Omega_{\q;0}}
\end{equation}
where
\begin{equation}\label{eq:correction-variance}
\delta_{\q} = -\frac{\mathfrak{Re}\Omega_{\q;1}}{\mathfrak{Re}\Omega_{\q;0}}-\frac{3\mathfrak{Re}\Gamma_{\q}}{2(\mathfrak{Re}\Omega_{\q;0})^2} \,.
\end{equation}
Furthermore, as we are interested in large scales, we will evaluate $\braket{v_{\q}^2}$ in the superhorizon limit $k\eta\to0^-$. Finally, we will see in what follows that $\mathfrak{Re}\Omega_{\q}$ and $\mathfrak{Re}\Gamma_{\q}$ only depend on $k = |\K|$. Thus, $\mathcal{P}_v(\q) \equiv \mathcal{P}_v(k)$. 

\subsubsection{Power spectra for scalar and tensor modes}
Following the discussion in~\cite{BKK-1,BKK-2}, we now define the power spectra for the scalar and tensor perturbations. In the scalar case, we define the variables
\begin{equation}
\zeta_{\K}:=\sqrt{\frac{3\kappa}{\epsilon}}\frac{v_{\K}^{(\mathrm{S})}}{a} \,,
\end{equation}
where $\epsilon$ is the slow-roll parameter, $\epsilon=1-\dot{\mathcal{H}}/\mathcal{H}^2$. The variables $\zeta_{\K}$ describe comoving curvature perturbations, which are related to the CMB temperature anisotropies. The $\zeta_{\K}$ perturbations are well-defined in quasi-de Sitter space, where the slow-roll parameter is small but non-vanishing. The power spectrum for $\zeta_{\K}$ is found to be
\begin{equation}\label{eq:PS-S}
\mathcal{P}_{\mathrm{S}}(k) := \frac{3\kappa}{\epsilon a^2}\mathcal{P}_v(k) = \frac{3\kappa}{\epsilon a^2}\frac{k^3}{2\pi^2}\braket{v_{\q}^2} .
\end{equation}
In the tensorial case, we consider the power spectrum for the variables $\sqrt{2}h_{\K}^{(+,\times)}$, where $h_{\K}^{(+,\times)}$ are the Fourier modes given in~\eref{tensor-fourier}. Summing over both polarizations, one obtains
\begin{equation}\label{eq:PS-T}
\mathcal{P}_{\mathrm{T}}(k) := \sum_{\lambda = +,\times}\frac{24\kappa}{a^2}\mathcal{P}_v(k) = \frac{48\kappa}{a^2}\frac{k^3}{2\pi^2}\braket{v_{\q}^2} .
\end{equation}
In the quasi-de Sitter case in which we are working, the power spectra for both the scalar and tensor perturbations are corrected by the same factor $\delta_{\q}$ given in~\eref{correction-variance}. This is a consequence of the equality of the frequencies~\eref{freq-deSitter} and no longer holds for general slow-rolls models~\cite{BKK-2} (see also the discussion in Sec.~\ref{sec:conclusions}). Indeed, due to~\eref{correction-variance-0}, we may rewrite~\eref{PS-S} and~\eref{PS-T} as
\begin{equation}\label{eq:correction-power-spectra}
\mathcal{P}_{\mathrm{S},\mathrm{T}}(k) = \mathcal{P}_{\mathrm{S},\mathrm{T};0}(k)(1+\kappa\delta_{\q}) \,.
\end{equation}
Moreover, the tensor-to-scalar ratio is defined as
\begin{equation}\label{eq:tensor-to-scalar}
r:=\frac{\mathcal{P}_{\mathrm{T}}(k)}{\mathcal{P}_{\mathrm{S}}(k)} \equiv \frac{\mathcal{P}_{\mathrm{T};0}(k)}{\mathcal{P}_{\mathrm{S};0}(k)} \,,
\end{equation}
and it is not corrected in the quasi-de Sitter case due to~\eref{correction-power-spectra}.

\subsubsection{Uncorrected power spectra}
As discussed in Sec.~\ref{sec:qft}, we recover the limit of QFT on a classical de Sitter background at order $\kappa^0$. Thus, we only reproduce known results at this order. Equation~\eref{Om} reads
\begin{equation}\label{eq:Om-schrodinger-kappa0}
\I\del{\Omega_{\q;0}}{\eta} = \Omega_{\q;0}^2-\omega_{\q}^2 \,.
\end{equation}
As is well-known, this can be solved via the substitution
\begin{equation}\label{eq:Om-y}
\Omega_{\q;0}(\eta) = -\I\frac{\dot{y}_{\q}(\eta)}{y_{\q}(\eta)} \,,
\end{equation}
which yields
\begin{equation}\label{eq:Om-schrodinger-kappa0-2}
\ddot{y}_{\q}+\omega_{\q}^2y_{\q} = 0 \,.
\end{equation}
Using~\eref{freq-deSitter}, one is able to find the well-known solution, $y_{\q}(\eta) \equiv y_{\K}(\eta)$,
\begin{equation}
y_{\K}(\eta) = \frac{A}{\sqrt{2k}}\e^{-\I k\eta}\left(1-\frac{\I}{k\eta}\right)+\frac{B}{\sqrt{2k}}\e^{\I k\eta}\left(1+\frac{\I}{k\eta}\right) .
\end{equation}
Due to~\eref{Om-y}, the normalization of $y_{\K}(\eta)$ is irrelevant to the computation of $\Omega_{\q;0}(\eta)$. Only the ratio of the constants $A,B$ is important. The requirement $\mathfrak{Re}\Omega_{\q;0}(\eta) > 0$ translates into the Wronskian condition
\begin{equation}
B^2-A^2 = -\I\squarebr{\dot{y}_{\K}y^{*}_{\K}-\dot{y}^{*}_{\K}y_{\K}} > 0 \,.
\end{equation}
This is fulfilled if $B\propto\cosh\vartheta$ and $A\propto\sinh\vartheta$. The (real) value of $\vartheta$ can be determined by imposing that~\eref{master-ansatz} reproduces the Minkowski vacuum at order $\kappa^0$ in the infinite past. Indeed, we find from~\eref{Om-y}
\begin{equation}
\Omega_{\K;0}(\eta) \stackrel{\eta\to-\infty}{\simeq}\frac{\cosh\vartheta\e^{\I k\eta}-\sinh\vartheta\e^{-\I k\eta}}{\cosh\vartheta\e^{\I k\eta}+\sinh\vartheta\e^{-\I k\eta}}k \,.
\end{equation}
We therefore require $\vartheta = 0$ to obtain the Minkowski vacuum. In this way, we recover the Bunch-Davies mode functions\footnote{In fact, the functions that appear in~\eref{Om-y} and~\eref{BD-modes} are the complex conjugates of the usual Heisenberg-picture mode functions~\cite{BKK-1}.}
\begin{equation}\label{eq:BD-modes}
y_{\K}(\eta)\propto \frac{1}{\sqrt{2k}}\e^{\I k\eta}\left(1+\frac{\I}{k\eta}\right) ,
\end{equation}
from which we find~\cite{BKK-1}
\begin{equation}\label{eq:Om-kappa-0}
\Omega_{\q;0}(\eta) \equiv \Omega_{k;0}(\eta) = \frac{k^3\eta^2}{1+k^2\eta^2}+\frac{\I}{\eta(1+k^2\eta^2)} \,,
\end{equation}
and, up to order $\kappa^0$,
\begin{equation}\label{eq:late-time-lowest-order-variance}
\braket{v_{\q}^2} = \frac{1+k^2\eta^2}{2k^3\eta^2}\stackrel{k\eta\to0^-}{\simeq}\frac{1}{2k^3\eta^2} \,.
\end{equation}
The corresponding power spectra have the well-known form [cf.~\eref{PS-S} and~\eref{PS-T}]
\begin{align}
\mathcal{P}_{\mathrm{S};0}(k) &= \frac{3\kappa}{\epsilon a^2}\frac{1}{4\pi^2\eta^2} = \left.\frac{GH_0^2}{\pi\epsilon}\right|_{k = aH_0} ,\label{eq:PS-S-deSitter}\\
\mathcal{P}_{\mathrm{T};0}(k) &=  \frac{24\kappa}{a^2}\frac{1}{4\pi^2\eta^2} = \frac{16GH_0^2}{\pi} \,,\label{eq:PS-T-deSitter}\\
r &=16\epsilon \,.
\end{align}
In~\eref{PS-S-deSitter} and~\eref{PS-T-deSitter}, we used $\kappa = 4\pi G/3$ and~\eref{conformal-time}. Moreover, due to the presence of $\epsilon$ in~\eref{PS-S-deSitter}, we evaluate the scalar power spectrum at the instant $k = aH_0$, which is justified since at order $\kappa^0$ the curvature perturbations freeze at horizon crossing.
\begin{widetext}
\subsubsection{Corrected power spectra}
As discussed in Sec.~\ref{sec:expansion} viz.~Sec.~\ref{sec:corrected-schrodinger}, we obtain corrections to the usual quantum-field-theoretical results on a de Sitter background at order $\kappa$ in the weak-coupling expansion. The corresponding effects on the power spectra can be found as follows.

First, we replace $\Omega_{\q}$ by its lowest-order value given in~\eref{Om-kappa-0} in~\eref{Ga} and solve this equation for $\Gamma_{\q}$. A boundary condition must be chosen and we choose to require that $\Gamma_{\q}$ vanishes in the infinite past. In this way, we find the solution
\begin{equation}\label{eq:Ga-sol}
\begin{aligned}
\Gamma_{\q}(\eta) &= \frac{H_0^2\eta\left(4 \I k^2\eta^2+4k\eta+\I\right)\e^{4 \I\arctan(k\eta)} }{6 \left(k^2\eta^2+1\right)^2}\\
& \;\;-\frac{8 H_0^2\eta^4 k^3 \Gamma(0,-4 \I k \eta)\e^{-4 \I \left[k\eta-\arctan(k\eta)\right]}}{3 \left(k^2\eta^2+1\right)^2} \,,
\end{aligned}
\end{equation}
where $\Gamma(0,z)$ is the upper incomplete gamma function.\footnote{As in~\cite{BKK-1}, we have used the upper incomplete gamma function $\Gamma(0,z)$ instead of the exponential integral function that may be presented in the solution to~\eref{Ga} by some computer algebra programs.}
Using
\begin{equation}\label{eq:gamma-past}
\Gamma(0,z) \stackrel{z\to-\infty}{\simeq} \frac{\e^{-z}}{z} \,,
\end{equation}
we find that the solution \eref{Ga-sol} obeys the chosen boundary condition
\begin{equation}\label{eq:Ga-past}
\lim_{\eta\to -\infty}\Gamma_{\q}(\eta) = 0 \,.
\end{equation}
Moreover, using
\begin{equation}\label{eq:gamma-future}
\Gamma(0,z) = -\gamma_E-\log z+z + \Ob(z^2) \,,
\end{equation}
where $\gamma_E$ is the Euler-Mascheroni constant, we find the late-time behavior of $\mathfrak{Re}\Gamma_{\q}$,
\begin{equation}\label{eq:Ga-future}
\mathfrak{Re}\Gamma_{\q}(\eta) \stackrel{\eta\to0^-}{\simeq} \frac{k^3H_0^2\eta^4}{3}\left[-18+8\gamma_E+8\log(4k|\eta|)\right] .
\end{equation}
The second step is to solve~\eref{Om} by using the expansion $\Omega_{\q} = \Omega_{\q;0}+\kappa\Omega_{\q;1}$. Up to order $\kappa$, we obtain
\begin{equation}\notag
\I\del{\Omega_{\q;1}}{\eta} = 2\Omega_{\q;0}\Omega_{\q;1}-3\Gamma_{\q}-\frac{3 H_0^2\eta^4}{2}\Omega_{\q;0}\left(\Omega_{\q;0}^2-\omega_{\q}^2\right) .
\end{equation}
Using~\eref{Om-kappa-0} and~\eref{Ga-sol}, we find the solution
\begin{equation}\label{eq:Om1}
\begin{aligned}
\Omega_{\q;1}(\eta) &= \frac{\e^{2 \I \arctan(k\eta)}H_0^2\eta^2}{k\eta+\I}\left[\frac{10\I+6 k\eta-3 \I k^2 \eta^2}{2(k\eta-\I) (k\eta+\I)}-\frac{4 \Gamma (0,-4 \I k \eta )}{(k\eta+\I)}\e^{-4 \I k\eta}-\frac{2 \Gamma (0,-2 \I k \eta)}{(k\eta-\I)}\e^{-2 \I k\eta}\right] ,
\end{aligned}
\end{equation}
\end{widetext}
where we have chosen the integration constant such that $\mathfrak{Re}\Omega_{\q;1}$ has a well-defined limit in the infinite past, i.e., such that it does not exhibit oscillatory behaviour in the limit $\eta\to-\infty$. A similar prescription was adopted in~\cite{BKK-1}. Indeed, using~\eref{gamma-past}, we obtain
\begin{equation}\label{eq:Om1-past}
\lim_{\eta\to-\infty}\mathfrak{Re}\Omega_{\q;1}(\eta) = \frac{3 H_0^2}{2 k^2} \,.
\end{equation}
We take~\eref{Ga-past} and~\eref{Om1-past} to be the order-$\kappa$ boundary conditions that are analogous to the usual order-$\kappa^0$ Bunch-Davies choice.\footnote{There is a large early-time contribution from $\mathfrak{Im}\Omega_{\q;1}$ to the phase of the wave function~\eref{master-ansatz}. However, this does not affect the conditional correlation functions of the $v_{\q}$ variables.} Using~\eref{gamma-future}, we also obtain the late-time behavior
\begin{equation}\label{eq:Om1-future}
\begin{aligned}
\mathfrak{Re}\Omega_{\q;1}(\eta) \stackrel{\eta\to0^-}{\simeq} H_0^2\eta^2\bigl[5 -2\gamma_E &+2\log(2k|\eta|) \\ &\;\;-4\log(4k|\eta|)\bigr]\,.
\end{aligned}
\end{equation}
From~\eref{late-time-lowest-order-variance},~\eref{Ga-future} and~\eref{Om1-future}, the correction~\eref{correction-variance} reads
\begin{equation}\label{eq:correction-result}
\delta_{\q}\equiv\delta_{k}(\eta) = H_0^2\left(\frac{k_{\star}}{k}\right)^3\bigl[4-2\gamma_E-2\log(-2k\eta)\bigr]\,,
\end{equation}
where we have inverted the redefinition of $k$ given in~\eref{Lfrak-redef-2}, i.e., we have rescaled $k\to\mathfrak{L}k$ and we have set $\mathfrak{L} = 1/k_{\star}$. In this way, the variable $k$ is again dimensionful and can be compared to observations, whereas $k_{\star}$ corresponds to a reference scale that can be identified with the pivot scale used in the CMB data analysis. Inserting the result \eref{correction-result} into the expression for the corrected variance \eref{correction-variance-0} and subsequently into the definition of the power spectra \eref{PS-S} and \eref{PS-T}, we obtain the following result for the corrected scalar and tensor power spectra: 
\begin{align}
\mathcal{P}^\text{corr}_{\mathrm{S,T}}&(k) = \mathcal{P}_{\mathrm{S,T};0}(k)\left[1 + \kappa\delta_{\q}\right] \\
\simeq\;&\mathcal{P}_{\mathrm{S,T};0}(k) \left[1 + \kappa H_0^2\left(\frac{k_{\star}}{k}\right)^3\bigl[2.85-2\log(-2k\eta)\bigr]\right]. \nonumber
\end{align}
This result differs from the one found in~\cite{BKK-1}\footnote{The rescaled Planck mass $m_\text{P}^2$ used in \cite{BKK-1} corresponds to $\kappa = m_\text{P}^{-2}$.}, most notably due to the presence of the logarithmic term. Let us now discuss the source and physical interpretation of the differences between~\eref{correction-result} and the result of~\cite{BKK-1}.

\subsubsection{\label{sec:discussion}Discussion}
Although the formalism presented here may be seen as a justification of why the would-be unitarity-violating terms may be neglected [they are absorbed into the measure $\hat{\mu}$ given in~\eref{mu-kappa2}], our result~\eref{correction-result} differs from the one presented earlier in~\cite{BKK-1} because other terms were neglected in that reference. Indeed, in addition to the terms that have been cancelled by the definition of $\hat{\mu}$, the imaginary part of the order-$\kappa$ terms in~\eref{Om} was also ignored in~\cite{BKK-1} (as well as in~\cite{BKK-2,BKK-3,Stein2}). The reason for this was that these terms were also considered to violate unitarity in these works. Here, however, we see that this is not the case. The effective Hamiltonian~\eref{H-eff} is self-adjoint provided suitable boundary conditions in field space are adopted, and we have explicitly verified that the norm of the conditional wave function is conserved [cf.~\eref{norm-is-conserved}]. Thus, by taking into account these additional terms, we have obtained a different result, mainly due to the appearance of the logarithmic term.

What is the significance of the logarithmic term in~\eref{correction-result}? We note that similar terms that grow logarithmically in conformal time often appear in perturbative QFT in de Sitter space, when one calculates perturbative corrections to correlators as well as the late-time Bunch-Davies wave function~\cite{Starobinsky:1982,Weinberg,DRG1,DRG2,Anninos}. The growing logarithms are an example of secular terms, which were first found in correlators for a massless scalar field in de Sitter space in~\cite{Starobinsky:1982}. That such growing logarithms should also appear in the present context is perhaps plausible for the following reason. The correction terms that stem from the weak-coupling expansion described here are analogous to loop corrections~\cite{Barvinsky:1997} in the sense that the master WDW equation~\eref{master-WDW-deSitter} takes into account the quantum nature of the background and, thus, corrects the usual quantum-field-theoretic description in which the background spacetime is classical. Hence, our late-time logarithmic growth originates from quantum-gravitational corrections, whereas the logarithms discussed in the literature~\cite{Starobinsky:1982,Weinberg,DRG1,DRG2,Anninos,DRG3,DRG4} arise from quantum corrections of perturbative QFT. It is also worthwhile to mention that a similar logarithmic term was found in the approach to the master WDW equation based on quantum moments that was presented in~\cite{Brizuela:moment}.

As this logarithm appears in the late-time superhorizon limit, it yields a potentially large contribution to the power spectra (it diverges as $\eta\to0^-$) which jeopardizes the validity of perturbation theory. While we have formally established the unitarity of the dynamics, this evidently rests on the assumption that the weak-coupling expansion is well-defined. Can we then ensure that this is the case? Let us consider two broad possibilities to be pursued in future work.

First, one may speculate that going beyond the de Sitter approximation would cure the weak-coupling corrections from the large logarithm. This is justified by considering the fact that the correction~\eref{correction-result} is valid for a quasi-de Sitter background and, in fact, slow-roll corrections need to be taken into account in realistic scenarios~\cite{BKK-2,BKK-3}, in which the structure of~\eref{correction-result} could be altered. Alternatively, one could engineer an ansatz for the conditional wave functions that is more involved than~\eref{master-ansatz} with the goal of avoiding such logarithms.

Second, one could find inspiration in the several ways in which large time-dependent logarithms have been addressed in the context of QFT in de Sitter space~\cite{DRG2}. In particular, it is conceivable that certain resummation techniques available in the literature~\cite{Starobinsky:1994}, such as the one based on the stochastic approach of~\cite{Starobinsky:1986}, could also be employed in the present context to remove the secular divergences. An appealing approach is the use of the dynamical renormalization group (DRG)~\cite{Chen,DRG1,DRG2,DRG3,DRG4}. In analogy to the conventional renormalization group, we define a subtraction procedure in the DRG framework by introducing an arbitrary time scale and by removing the late-time divergences with the inclusion of appropriate counterterms~\cite{Chen}. The DRG techniques allow us to, in principle, resum the leading time-dependent logarithms and improve perturbation theory. Whether this DRG improvement can be applied to the weak-coupling expansion of the BO approach is an interesting topic that could be examined in the future. It is also worth mentioning that in the recent article~\cite{Kamenshchik:2020} a technique of resummation of secular terms inspired by the renormalization group was applied, with results similar to those of the stochastic approach~\cite{Starobinsky:1994,Starobinsky:1986}.

We note, however, that a rough physical interpretation and estimates can be given to~\eref{correction-result}. If one invokes the conservation of the comoving curvature perturbation on superhorizon scales, one could evaluate~\eref{correction-result} near horizon crossing [when $\log(-k\eta)\simeq 0$]. In that case, one finds scale dependence and an enhancement of power on the largest scales,
\begin{equation}
\kappa\delta_{k} \simeq 1.5\,\kappa H_0^2\left(\frac{k_{\star}}{k}\right)^3 , 
\end{equation}
which is similar to the result of~\cite{BKK-1}, where the numerical factor in this term was determined to be approximately 0.988. More generally, one notices that $\log(-k\eta)$ is proportional to the number of e-folds elapsed between the time at which the mode $k$ crosses the horizon (given by $aH_0 = k$) and the time $\eta$ at which the correction term is computed~\cite{DRG1,DRG2}. Thus, one could evaluate~\eref{correction-result} at a certain number of e-folds for which perturbation theory would still be valid, i.e., for which $\kappa\delta_{k}(\eta)$ would be small, taking into account the upper bound $\kappa H_0^2\lesssim 1.7\times 10^{-10}$ computed in~\cite{BKK-1}. For example, one could consider that the logarithm is of the order of $60$ e-folds ($|\log(-k\eta)|\lesssim60$). To determine whether such estimates are reasonable, one needs a better physical understanding of the weak-coupling expansion of the master WDW equation for more realistic (slow-roll) models.

\section{\label{sec:conclusions}Conclusions}
The weak-coupling expansion of the BO approach to quantum gravity has been frequently used~\cite{LapRuba:1979,Banks:1984,Halliwell:1984,Brout:1987-4,PadSingh:1990-1,PadSingh:1990-2,Kiefer:1993-2,Parentani:1997-2,Kiefer:1991,Singh:1990,Kiefer:1991,Bertoni:1996,Bologna:2017,Kiefer:2018,Chataig:2019-1} to recover QFT in curved spacetimes as a limit of quantum gravity as well as to obtain corrections to the dynamics of quantum fields on curved backgrounds. One of the main applications of this approach has been the calculation of corrections to primordial power spectra in inflation~\cite{BKK-0-0,BKK-0-1,BKK-1,BKK-2,BKK-3,Stein2}. Although the corrections to the power spectra are typically found to be small and are currently unobservable~\cite{BKK-1,BKK-2}, it is an exciting prospect that these corrections may lead to indirect effects in galaxy-galaxy correlation functions and other phenomena related to structure formation, which might be observable in future measurements. However, to ascertain whether this hope is grounded, one must first verify if the weak-coupling expansion is well-defined and the ensuing theory is consistent.

A major source of controversy has been the issue of the unitarity of the BO approach. Whereas some works~\cite{Kiefer:1991,BKK-0-0,BKK-0-1,BKK-1,BKK-2,BKK-3,Stein2} argued for the existence of unitarity-violating terms, others~\cite{Bertoni:1996,Bologna:2017,Kiefer:2018} argued that the inclusion of backreaction terms would ensure that the norm of the states was preserved. Following~\cite{Chataig:2019-1}, we have argued that both approaches are equivalent and that the backreaction terms are a priori ambiguous. We have noted that one may secure unitarity by redefining the wave function in such a way that it is normalized at all times and that this corresponds to a fixation of the backreaction terms. However, this is not a preferred procedure, since it depends \emph{non-linearly} on the state at hand [cf.~discussion in Sec.~\ref{sec:ambiguity-backreaction}, in particular,~\eref{enforcing-unitarity-gamma}]. In contrast, we have shown that perturbative unitarity is obtained, provided one reinterprets the would-be unitarity-violating terms as precisely those that define the inner product measure [cf.~\eref{mu-kappa}]. This measure is a \emph{linear} operator and its definition is not \emph{ad hoc}, as it arises naturally in the weak-coupling expansion.

Furthermore, one of the principal features of the BO approach is the definition of a background time variable, conditioned on which the dynamics of cosmological perturbations unfolds. As the system of background degrees of freedom and cosmological perturbations is assumed to be time-reparametrization invariant, the definition of the background time variable should correspond in some sense to a ``gauge choice'' with an associated Faddeev-Popov determinant. Indeed, we have shown in Sec.~\ref{sec:relation-gf} that the perturbative measure is related to a specific notion of gauge fixing the time variable in quantum cosmology. The positive-definite, perturbative inner product can be rewritten as a matrix element of an operator, the classical limit of which is the Faddeev-Popov determinant [cf.~\eref{conditional-FP}]. The perturbative inner product thus coincides with a gauge-fixed inner product.

In this way, the present article can be seen as a step in the direction of unifying the BO approach with recent developments in relational approaches to quantum dynamics~\cite{Chataig:2019-2,Hoehn:2018-1,Hoehn:2018-2,Hoehn:2019,Hoehn:Trinity,Chataig:2020,Hoehn:2020}. The main idea is that the time-reparametrization-invariant dynamics can be expressed in terms of relational quantities, which are defined in the quantum theory through the gauge-fixed inner product and its associated conditional probabilities [cf. Sec.~\ref{sec:quantum-flrw}]. Our results indicate that the BO approach may be seen as a particular case of a more general relational formalism. Although it is at present unclear whether such a relational theory describes nature at the fundamental level, we believe it is worthwhile to explore the possible phenomenological consequences of this paradigm.

There are several possible continuations of the present article. First, one can generalize our results regarding unitarity and the calculation of corrections to the primordial power spectra to general slow-roll models~\cite{BKK-2}. Indeed, we have restricted ourselves to the de Sitter background for simplicity and to illustrate the issue of unitarity in the simplest possible case, without the need to complicate the analysis with the inclusion of the slow-roll parameters. Nevertheless, the analysis of the more realistic slow-roll models may clarify the physical interpretation of the correction terms, in particular of the logarithmic term found in~\eref{correction-result}. Second, as was discussed in Sec.~\ref{sec:discussion}, there may be interesting ways to improve the validity of perturbation theory, e.g., by resumming the large logarithmic contributions using the DRG. It may also be that a proper treatment of the large logarithm leads to an enhancement of the size of the correction terms and their observability. Third, a crucial element of the calculation of the correction terms is the random phase approximation [cf.~Sec.~\ref{sec:single-mode}]. It would be interesting to verify the conditions under which this approximation holds by adopting a certain regularization scheme. These stimulating sequels would put the BO approach on firmer ground and pave the way to well-defined and possibly observable quantum-gravitational effects from the early Universe.

\begin{acknowledgments}
The authors thank David Brizuela, Claus Kiefer, and Branislav Nikoli\'{c} for useful discussions, and an anonymous referee for constructive comments. L.\,C.~gratefully acknowledges financial support from the Bonn-Cologne Graduate School of Physics and Astronomy. L.\,C.~also thanks the Institute for Theoretical Physics at KU Leuven for kind hospitality while part of this research was done. The work of M.\,K.~was supported in part by the KU Leuven C1 grant ZKD1118 C16/16/005 and the European Research Council Grant No.~ERC-2013-CoG 616732 HoloQosmos.
\end{acknowledgments}

\end{document}